\documentclass{article}

\usepackage[nonatbib,final]{neurips_2024}

\usepackage[utf8]{inputenc} 
\usepackage[T1]{fontenc}    
\usepackage{hyperref}       
\usepackage{url}            
\usepackage{booktabs}       
\usepackage{amsfonts}       
\usepackage{nicefrac}       
\usepackage{microtype}      
\usepackage{xcolor}         

\usepackage[pdftex]{graphicx}

\usepackage{amsmath, amsthm, amssymb, bm}
\newtheorem{myDefinition}{Definition}
\newtheorem{myTheorem}{Theorem}
\newtheorem{myCorollary}{Corollary}
\newtheorem{myProposition}{Proposition}

\usepackage[backend=bibtex, style=authoryear,natbib=true]{biblatex}

\bibliography{_recor}

\title{Adjust Pearson's $r$ to Measure Arbitrary Monotone Dependence}

\author{%
Xinbo Ai \\
School of Intelligent Engineering and Automation\\
  Beijing University of Posts and Telecommunications\\
  Beijing 100876, China \\
  \texttt{axb@bupt.edu.cn} \\
}

\begin{document}

\maketitle

\begin{abstract}
  Pearson's $r$, the most widely-used correlation coefficient, is traditionally regarded as exclusively capturing linear dependence, leading to its discouragement in contexts involving nonlinear relationships. However, recent research challenges this notion, suggesting that Pearson's $r$ should not be ruled out \textit{a priori} for measuring nonlinear monotone relationships. Pearson's $r$ is essentially a scaled covariance, rooted in the renowned Cauchy-Schwarz Inequality. Our findings reveal that different scaling bounds yield coefficients with different capture ranges, and interestingly, tighter bounds actually expand these ranges. We derive a tighter inequality than Cauchy-Schwarz Inequality, leverage it to refine Pearson's $r$, and propose a new correlation coefficient, i.e., rearrangement correlation. This coefficient is able to capture arbitrary monotone relationships, both linear and nonlinear ones. It reverts to Pearson's $r$ in linear scenarios. Simulation experiments and real-life investigations show that the rearrangement correlation is more accurate in measuring nonlinear monotone dependence than the three classical correlation coefficients, and other recently proposed dependence measures.
\end{abstract}

\section{Introduction}

Proposed in the late 19th century, Pearson's $r$ \citep{pearsonMathematicalContributionsTheory1896} has been one of the main tools for scientists and engineers to study bivariate dependence during the 20th century. It is remarkably unaffected by the passage of time \citep{leerodgersThirteenWaysLook1988} and still goes strong in the 21st century \citep{puccettiMeasuringLinearCorrelation2022}. It has been, and probably still is, the most used measure for statistical associations, and generally accepted as \textit{the} measure of dependence, not only in statistics, but also in most applications of natural and social sciences \citep{tjostheimStatisticalDependencePearson2022}.

Despite its popularity, Pearson's $r$ has a number of shortcomings, and the most serious issue might be that it can only capture linear dependence, as stated in classical textbooks \citep{wassermanAllStatisticsConcise2004} and contemporary literatures \citep{armstrongShouldPearsonCorrelation2019,tjostheimStatisticalDependencePearson2022}. The use of Pearson's $r$ has been strongly discouraged for forms of associations other than linear ones \citep{speedCorrelation21stCentury2011}. 

Numerous nonlinear alternative coefficients have been proposed to address this deficiency, such as Spearman's $\rho$ \citep{spearmanProofMeasurementAssociation1904}, 
Kendall's $\tau$ \citep{kendallNEWMEASURERANK1938}, Hilbert-Schmidt Independence Criterion(HSIC) \citep{grettonMeasuringStatisticalDependence2005}, distance correlation(dCor) \citep{szekelyMeasuringTestingDependence2007}, Maximal Information Coefficient(MIC) \citep{reshefDetectingNovelAssociations2011}, and Chatterjee's $\xi$ \citep{chatterjeeNewCoefficientCorrelation2021}. Their capture ranges are extending from linear dependence to monotone dependence, and then to non-monotone dependence. Without exception, all these coefficients adopt radically different approaches for nonlinear dependence, rather than following the original way of Pearson's $r$ for a breakthrough.

In their recent paper titled ``\textit{Myths About Linear and Monotonic Associations: Pearson's $r$, Spearman's $\rho$, and Kendall's $\tau$}'', \citet{vandenheuvelMythsLinearMonotonic2022} challenged the widespread belief that Pearson's $r$ is only a measure for linear dependence,  proving this notion to be false. Their findings indicate that Pearson's $r$ should not be ruled out \textit{a priori} for measuring nonlinear monotone dependence. Although this potential has been recognized, the specific approach to using Pearson's $r$ for accurate measurement of nonlinear monotone dependence remains unresolved.

Pearson's $r$ is essentially a scaled covariance, with the renowned Cauchy-Schwarz Inequality as its mathematical foundation. We find that different scaling bounds yield coefficients with different capture ranges, and interestingly, tighter bounds actually expand these ranges. We derive a tighter inequality than Cauchy-Schwarz Inequality, leverage it to adjust Pearson's $r$ to measure nonlinear monotone dependence. The adjusted version of Pearson's $r$ is more accurate in measuring nonlinear monotone dependence than the three classical correlation coefficients, and other recently proposed dependence measures.

\section{Methods}
\subsection{Definitions and notations}

Consider real-valued random variables $X$ and $Y$ with cdf's (cumulative distribution functions) $F$ and $G$ respectively. We denote the covariance of $X$ and $Y$ as $\operatorname{cov} \left( {X,Y} \right)$; the variance of $X$ as $\operatorname{var} \left( X \right)$, and the variance of $Y$ as $\operatorname{var} \left( Y \right)$. We assume that $0 < \operatorname{var} \left( X \right) < \infty $, $0 < \operatorname{var} \left( Y \right) < \infty $. We define ${X^ \uparrow } = {F^{ - 1}}\left( U \right)$, ${X^ \downarrow } = {F^{ - 1}}\left( {1 - U} \right)$. Here, $U$ is a random variable with the uniform distribution on $\left( {0,1} \right)$, and ${F^{ - 1}}$ is the inverse cdf or quantile function defined as ${F^{ - 1}}\left( u \right) = \inf \left\{ {x \in \mathbb{R}:F\left( x \right) \geqslant u} \right\},u \in \left( {0,1} \right)$. Similarly, ${Y^ \uparrow } = {G^{ - 1}}\left( U \right)$, ${Y^ \downarrow } = {G^{ - 1}}\left( {1 - U} \right)$.

Let $x = \left( {{x_1}, \cdots ,{x_n}} \right)$ and $y = \left( {{y_1}, \cdots ,{y_n}} \right)$ be samples of $X$ and $Y$, each with $n$ elements. Neither $x$ nor $y$ is constant. We denote the sample covariance of $x$ and $y$ as ${s_{x,y}}$; the sample variance of $x$ as $s_x^2$; the sample variance of $y$ as $s_y^2$. We define the increasing and decreasing rearrangement of $x$ as ${x^ \uparrow } = \left( {{x_{\left( 1 \right)}},{x_{\left( 2 \right)}}, \cdots ,{x_{\left( n \right)}}} \right)$ and ${x^ \downarrow } = \left( {{x_{\left( n \right)}},{x_{\left( {n - 1} \right)}}, \cdots ,{x_{\left( 1 \right)}}} \right)$ respectively, with ${x_{\left( 1 \right)}} \leqslant {x_{\left( 2 \right)}} \leqslant  \cdots  \leqslant {x_{\left( n \right)}}$. Similarly, we define ${y^ \uparrow } = \left( {{y_{\left( 1 \right)}},{y_{\left( 2 \right)}}, \cdots ,{y_{\left( n \right)}}} \right)$, ${y^ \downarrow } = \left( {{y_{\left( n \right)}},{y_{\left( {n - 1} \right)}}, \cdots ,{y_{\left( 1 \right)}}} \right)$.

\begin{myDefinition}\label{definition01}
  A subset $S$ of ${\mathbb{R}^2}$ is non-decreasing (\textit{resp.} non-increasing) if and only if for all $\left( {{x_1},{y_1}} \right)$, $\left( {{x_2},{y_2}} \right)$ in $S$, ${x_1} < {x_2}$ implies ${y_1} \leqslant {y_2}$ (\textit{resp.} ${x_1} < {x_2}$ implies ${y_1} \geqslant {y_2}$). Random variables $X$ and $Y$ are called increasing (\textit{resp.} decreasing) monotone dependent if  $\left( {X,Y} \right)$ lies almost surely in a non-decreasing (\textit{resp.} non-increasing) subset of ${\mathbb{R}^2}$. Samples $x$ and $y$ are called increasing (\textit{resp.} decreasing) monotone dependent if $\left\{ {\left( {x,y} \right)} \right\}$ is  a non-decreasing (\textit{resp.} non-increasing) subset of ${\mathbb{R}^2}$.
\end{myDefinition}

Definition \ref{definition01} is sourced from \citep{mikusinskiFrechetBoundsRevisted1991}. Clearly Definition \ref{definition01} is symmetrical with respect to $X$ and $Y$. The monotone dependence outlined here encompasses a broader scope than definitions like the one in \citep{kimeldorfMonotoneDependence1978}, where ``\textit{each of $X$ and $Y$ is almost surely a monotone function of the other}''. This is primarily because it doesn't necessitate a one-to-one mapping.  Also, linear dependence, i.e., $P\left( {Y = \alpha X + \beta } \right) = 1$ at the population level or $y = a{x} + b$ at the sample level, is special case of monotone dependence, and we will refer to dependence which is monotone but not linear as ``\textit{nonlinear monotone dependence}''.

\subsection{Different bounds lead to different capture ranges}\label{sec2.1}
With Cauchy-Schwarz Inequality, the well-known covariance inequality can be directly derived as 
\begin{equation*}
\left| {{\mathop{\rm cov}} \left( {X,Y} \right)} \right| \leqslant \sqrt {{\mathop{\rm var}} \left( X \right){\mathop{\rm var}} \left( Y \right)},
\end{equation*}
thus the geometric mean of ${\mathop{\rm var}} \left( X \right)$ and ${\mathop{\rm var}} \left( Y \right)$, i.e., $\sqrt {\operatorname{var} \left( X \right)\operatorname{var} \left( Y \right)} $, mathematically provides a bound for covariance ${\mathop{\rm cov}} \left( {X,Y} \right)$, which ensures that Pearson's Correlation Coefficient
\begin{equation*}
  r \left( {X,Y} \right) = \frac{{{\text{cov}}\left( {X,Y} \right)}}{{\sqrt {{\text{var}}\left( X \right){\text{var}}\left( Y \right)} }}
\end{equation*}
always falls into the range $[ - 1, + 1]$. Scaled by $\sqrt {{\mathop{\rm var}} \left( X \right){\mathop{\rm var}} \left( Y \right)} $, Pearson's $r$ turns into a \textit{normalized covariance}, which is dimensionless and bounded. It possesses significant advantage over the original \textit{covariance} in the sense that its value will not be affected by the change in the units of $X$ and $Y$. 

A crucial issue that has been neglected so far is that boundedness doesn't ensure optimum. Scaling ${\mathop{\rm cov}} \left( {X,Y} \right)$ to the range $[ - 1, + 1]$ is not the only thing that matters. In fact, different bounds can be utilized to scale covariance to be bounded coefficients, as reported in previous works \citep{linConcordanceCorrelationCoefficient1989,zegersFamilyChancecorrectedAssociation1986}. 

For example, with the Mean Inequality Series \citep{hardyInequalities1952}, it is immediate that
\begin{equation*}
  \left| {{\text{cov}}\left( {X,Y} \right)} \right| \leqslant \sqrt {{\text{var}}\left( X \right){\text{var}}\left( Y \right)}  \leqslant \frac{1}{2}\left( {{\text{var}}\left( X \right) + {\text{var}}\left( Y \right)} \right)
\end{equation*}
in the sense that geometric mean $\sqrt {{\text{var}}\left( X \right){\text{var}}\left( Y \right)} $ is always less than or equal to arithmetic mean $\tfrac{1}{2}\left( {{\text{var}}\left( X \right) + {\text{var}}\left( Y \right)} \right)$ for nonnegative values ${{\text{var}}\left( X \right)}$ and ${{\text{var}}\left( Y \right)}$. Then we get a looser bound for covariance, i.e., ${\textstyle{1 \over 2}}\left( {{\mathop{\rm var}} \left( X \right) + {\mathop{\rm var}} \left( Y \right)} \right)$, with which another measure can be defined as follows\citep{zegersFamilyChancecorrectedAssociation1986}:
\begin{equation*}
  {r ^ + }\left( {X,Y} \right) = \frac{{{\text{cov}}\left( {X,Y} \right)}}{{\frac{1}{2}\left( {{\text{var}}\left( X \right) + {\text{var}}\left( Y \right)} \right)}}
\end{equation*}
${r ^ + }\left( {X,Y} \right)$ is named early as \textit{Additivity Coefficient} \citep{zegersFamilyChancecorrectedAssociation1986} and later as \textit{Standardized Covariance} \citep{andraszewiczStandardizedCovarianceMeasure2014}. It is proved that the capture range of  ${r ^ + }\left( {X,Y} \right)$ is limited to additive relationships, i.e., $Y =  \pm X + \beta$, which are special cases of linear relationships, i.e., $Y = \alpha{X} + \beta$, with $\alpha$ being fixed to $ \pm 1$ \citep{zegersFamilyChancecorrectedAssociation1986}. 

Further, we can find an even looser bound for covariance, in the sense that
\begin{equation*}
  \frac{1}{2}\left( {{\text{var}}\left( X \right) + {\text{var}}\left( Y \right)} \right) \leqslant \frac{1}{2}\left( {{\text{var}}\left( X \right) + {\text{var}}\left( Y \right) + {{\left| {{\bar X} - {\bar Y}} \right|}^2}} \right),  
\end{equation*}
and define a new measure as follows:
\begin{equation*}
  {r ^ = }\left( {X,Y} \right) = \frac{{{\text{cov}}\left( {X,Y} \right)}}{{\frac{1}{2}\left( {{\text{var}}\left( X \right) + {\text{var}}\left( Y \right) + {{\left| {{\bar X} - {\bar Y}} \right|}^2}} \right)}}
\end{equation*}
${r ^ = }$ is named as \textit{Concordance Correlation Coefficient} \citep{linConcordanceCorrelationCoefficient1989}, and it is designed to measure identical relationship, i.e., $Y =  \pm X$. When $X$ and $Y$ are both positive, it can be utilized to evaluate their agreement by measuring the variation from the 45° line through the origin \citep{linConcordanceCorrelationCoefficient1989}.

As for the abovementioned three measures, $r$, ${r^ + }$, and ${r^ = }$, they share the same numerator, ${\mathop{\rm cov}} \left( {X,Y} \right)$, the differences lie in their denominators. These denominators serve as bounds for $\operatorname{cov} \left( {X,Y} \right)$. Different bounds lead to different capture ranges. With the bounds being looser, their capture ranges are shrinking from linear ($Y = \alpha X + \beta$) towards additive ($Y =  \pm X + \beta$) and ultimately to identical ($Y =  \pm X$) relationships. The looser the bound, the narrower the capture range.

Up until now, all the efforts have only led to looser bounds and measures with narrower capture ranges. \textit{Could we possibly explore breakthroughs by approaching the problem from the opposite direction, aiming to achieve a tighter bound and consequently, devise a new measure with a broader capture range?}

The bound in Pearson's $r$ is intrinsically provided by Cauchy-Schwarz Inequality, which is one of the most widespread and useful inequalities in mathematics. Cauchy-Schwartz Inequality is so classic and reliable that one seldom tries to improve it. Both bounds in ${r^ + }$ and ${r^ = }$ are looser than that provided by Cauchy-Schwartz Inequality. To loosen Cauchy-Schwartz Inequality might be easy while to tighten such a classic inequality might be relatively difficult. However, we find that it is not impossible to improve the tightness of Cauchy-Schwarz Inequality. In other words, there exists sharper bound for covariance, which will be depicted in the next section.

\subsection{New inequality tighter than Cauchy-Schwarz Inequality}

Before deriving the new inequality, we will briefly review the classic Cauchy-Schwarz inequality, which is common in textbooks. The Cauchy–Schwarz inequality states that for $x$ and $y$, we have
\begin{equation*}
  \left| {\left\langle {x,y} \right\rangle } \right| \leqslant \left\| x \right\|\left\| y \right\|,
\end{equation*}
where $\left\langle { \cdot , \cdot } \right\rangle $ is the inner product. and $\left\|  \cdot  \right\|$ is the norm. The equality holds if and only if $x$ and $y$ are linearly dependent, i.e., $y = a{x}$ for some constant $a$. 

After defining an inner product on the set of random variables using the expectation of their product, i.e., $\left\langle {X,Y} \right\rangle  = EXY$, the Cauchy–Schwarz inequality becomes
\begin{equation*}
  \left| {EXY} \right| \leqslant \sqrt {E{X^2}E{Y^2}}.
\end{equation*}

Now, we will sharpen the Cauchy–Schwarz inequality. On the basis of the rearrangement theorems \citep{hardyInequalities1952}, we derive 6 theorems(corollaries/propositions) as follows.
\begin{myTheorem}\label{myTheorem01}
  For random variables $X$ and $Y$, we have
  \begin{equation*}
    \left| {EXY} \right| \leqslant \left| {E{X^ \uparrow }{Y^ \updownarrow }} \right| \leqslant \sqrt {E{X^2}E{Y^2}}. 
  \end{equation*}
  The equality on the left holds if and only if $X$ and $Y$ are monotone dependent, and the equality on the right holds if and only if $Y\mathop  = \limits^d \alpha X$, with $\operatorname{sgn} \left( {EXY} \right) = \operatorname{sgn} \left( \alpha  \right)$.
  
  Here, $\mathop  = \limits^d $ denotes equality in distribution, and $E{X^ \uparrow }{Y^ \updownarrow }$ is defined as:
  \begin{equation*}
    E{X^ \uparrow }{Y^ \updownarrow } = \left\{ {\begin{array}{*{20}{c}}
      {E{X^ \uparrow }{Y^ \uparrow },if\;\;EXY\geqslant0} \\ 
      {E{X^ \uparrow }{Y^ \downarrow },if\;\;EXY < 0} 
    \end{array}} \right.    
  \end{equation*}
\end{myTheorem}

For the sake of conciseness, the proofs of Theorem \ref{myTheorem01} and theorems undermentioned are all included in Appendix \ref{appendixProof}.

\begin{myTheorem}\label{myTheorem02}
  For samples $x$ and $y$ we have
    \begin{equation*}
    \left| {\left\langle {x,y} \right\rangle } \right| \leqslant \left| {\left\langle {{x^ \uparrow },{y^ \updownarrow }} \right\rangle } \right| \leqslant \left\| x \right\|\left\| y \right\|.
  \end{equation*}
  The equality on the left holds if and only if $x$ and $y$ are monotone dependent, and the equality on the right holds if and only if $y$ is arbitrary permutation of $a{x}$, with $\operatorname{sgn} \left( {\left\langle {x,y} \right\rangle } \right) = \operatorname{sgn} \left( a \right)$. 
  
  Here, $\left\langle {{x^ \uparrow },{y^ \updownarrow }} \right\rangle $ is defined as:
  \begin{equation*}
    \left\langle {{x^ \uparrow },{y^ \updownarrow }} \right\rangle  = \left\{ {\begin{array}{*{20}{c}}
      {\left\langle {{x^ \uparrow },{y^ \uparrow }} \right\rangle ,if\;\;\left\langle {x,y} \right\rangle  \geqslant 0} \\ 
      {\left\langle {{x^ \uparrow },{y^ \downarrow }} \right\rangle ,if\;\;\left\langle {x,y} \right\rangle  < 0} 
    \end{array}} \right.
  \end{equation*}
\end{myTheorem}

\begin{myCorollary}\label{myCorollary01}
  For random variables $X$ and $Y$, we have covariance equality series as:
  \begin{equation*}
    \begin{aligned}
      \left| {\operatorname{cov} \left( {X,Y} \right)} \right| \leqslant \left| {\operatorname{cov} \left( {{X^ \uparrow },{Y^ \updownarrow }} \right)} \right| &\leqslant \sqrt {\operatorname{var} \left( X \right)\operatorname{var} \left( Y \right)}  \\
      &\leqslant \tfrac{1}{2}\left( {\operatorname{var} \left( X \right) + \operatorname{var} \left( Y \right)} \right) \\
      &\leqslant \tfrac{1}{2}\left( {\operatorname{var} \left( X \right) + \operatorname{var} \left( Y \right) + {{\left| {{\bar X} - {\bar Y}} \right|}^2}} \right)      
    \end{aligned}
  \end{equation*}
  The first equality holds if and only if $X$ and $Y$ are monotone dependent, and the second equality holds if and only if $Y\mathop  = \limits^d \alpha X + \beta $, with $\operatorname{sgn} \left( {\operatorname{cov} \left( {X,Y} \right)} \right) = \operatorname{sgn} \left( \alpha  \right)$.

  Here, $\operatorname{cov} \left( {{X^ \uparrow },{Y^ \updownarrow }} \right)$ is defined as:
  \begin{equation*}
    {\text{cov}}\left( {{X^ \uparrow },{Y^ \updownarrow }} \right) = \left\{ {\begin{array}{*{20}{c}}
      {{\text{cov}}\left( {{X^ \uparrow },{Y^ \uparrow }} \right),\;\;if\;\;{\text{cov}}\left( {X,Y} \right) \geqslant 0} \\ 
      {{\text{cov}}\left( {{X^ \uparrow },{Y^ \downarrow }} \right)\;\;if\;\;{\text{cov}}\left( {X,Y} \right) < 0} 
    \end{array}} \right.
  \end{equation*}
\end{myCorollary}

\begin{myCorollary}\label{myCorollary02}
  For samples $x$ and $y$, we have covariance inequality series as
  \begin{equation*}
    \begin{aligned}
    \left| {{s_{x,y}}} \right| \leqslant \left| {{s_{{x^ \uparrow },{y^ \updownarrow }}}} \right| &\leqslant \sqrt {s_x^2s_y^2}  \\
    &\leqslant \frac{1}{2}\left( {s_x^2 + s_y^2} \right) \\
    &\leqslant \frac{1}{2}\left( {s_x^2 + s_y^2 + {{\left| {\bar x - \bar y} \right|}^2}} \right)
  \end{aligned}
  \end{equation*}
  The first equality holds if and only if $x$ and $y$ are monotone dependent, and the second equality holds if and only if $y$ is arbitrary permutation of $a{x} + b$, with $\operatorname{sgn} \left( {{s_{x,y}}} \right) = \operatorname{sgn} \left( a \right)$. 

  Here, ${s_{{x^ \uparrow },{y^ \updownarrow }}}$ is defined as:
  \begin{equation*}
    {s_{{x^ \uparrow },{y^ \updownarrow }}} = \left\{ {\begin{array}{*{20}{c}}
      {{s_{{x^ \uparrow },{y^ \uparrow }}},if\;\;{s_{x,y}} \geqslant 0} \\ 
      {{s_{{x^ \uparrow },{y^ \downarrow }}},if\;\;{s_{x,y}} < 0} 
    \end{array}} \right.
  \end{equation*}
\end{myCorollary}

\subsection{The proposed Rearrangement Correlation}

The inequality series in Corollary \ref{myCorollary01} and Corollary \ref{myCorollary02} provides sharper bounds for covariance at the population level and the sample level respectively. We will leverage them to define the so-called \textit{Rearrangement Correlation}, which is the adjusted version of Pearson's $r$ proposed here.

\begin{myDefinition}\label{definition02}
  The Rearrangement Correlation of random variables $X$ and $Y$ is defined as:
  \begin{equation*}
    {r ^\# }\left( {X,Y} \right) = \frac{{\operatorname{cov} \left( {X,Y} \right)}}{{\left| {\operatorname{cov} \left( {{X^ \uparrow },{Y^ \updownarrow }} \right)} \right|}}
  \end{equation*}
\end{myDefinition}
\begin{myDefinition}
  The Rearrangement Correlation of samples $x$ and $y$ is defined as:
  \begin{equation*}
    {r^\# }\left( {x,y} \right) = \frac{{{s_{x,y}}}}{{\left| {{s_{{x^ \uparrow },{y^ \updownarrow }}}} \right|}}
  \end{equation*}
\end{myDefinition}

\begin{figure*}[!b]
  \centering
  \includegraphics[width=0.8\linewidth]{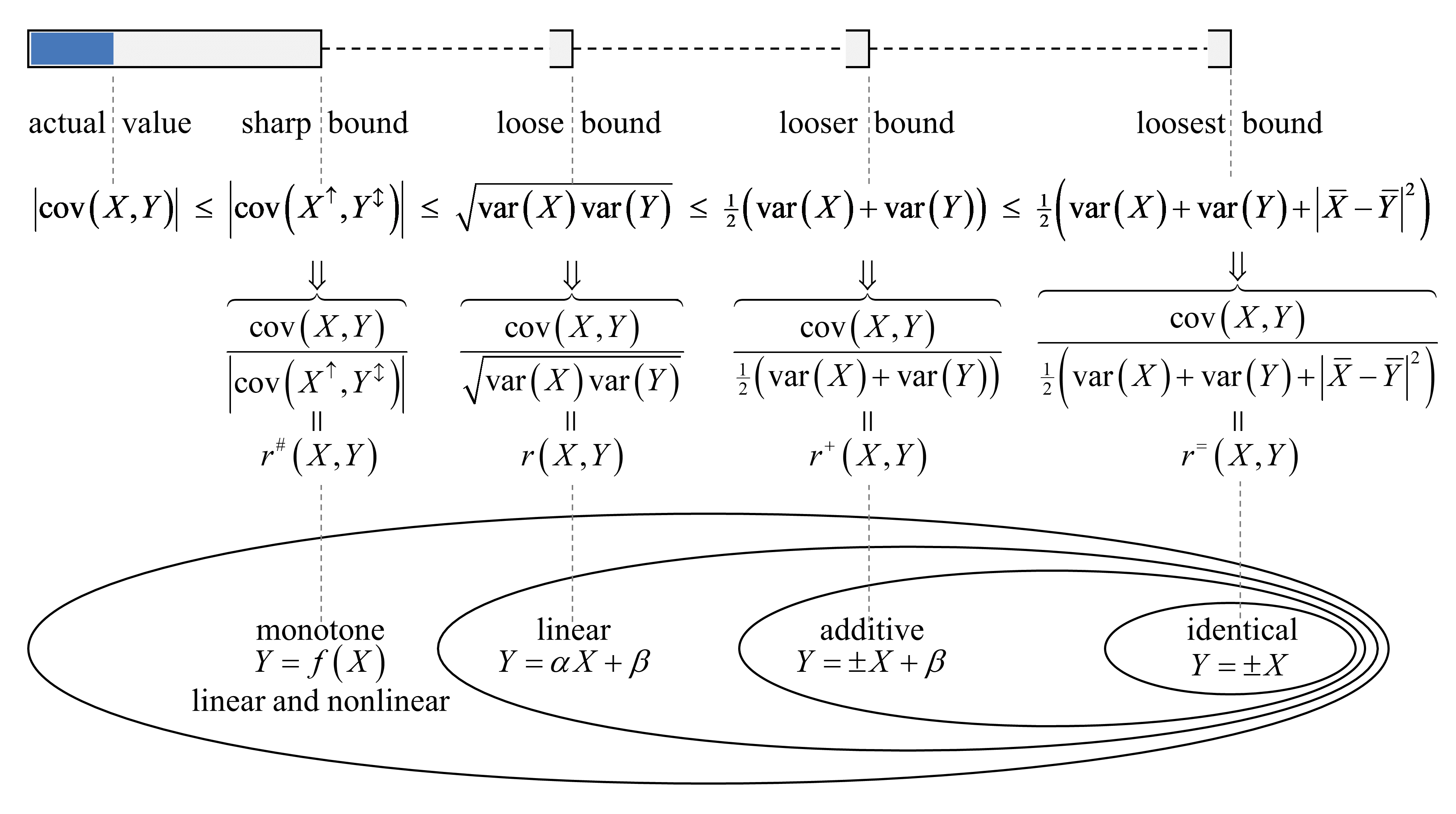}
  \caption{Covariance inequality series, correlation coefficients and their capture ranges}
  \label{fig1}
\end{figure*}

The new measure is named ``\textit{Rearrangement Correlation}'' because its theoretical foundation is the rearrangement inequality, as shown in Theorem \ref{myTheorem01} and Theorem \ref{myTheorem02}. We adopt the musical sharp symbol \# to denote rearrangement correlation, signifying that this measure has sharp values because of its sharp bounds. Analogous to how C\# is pronounced as \textit{C-sharp}, ${r^\# }$ is pronounced as \textit{r-sharp}. 

As for the relationship between ${r^\# }\left( {x,y} \right)$ and ${r ^\# }\left( {X, Y} \right)$, it is clear that ${r^\# }\left( {x,y} \right)$ converges to ${r ^\# }\left( {X,Y} \right)$ when $n \to \infty $ according to their definitions.

The capture range of rearrangement correlation is no longer limited to linear dependence but monotone dependence, which is revealed by the next proposition.
\begin{myProposition}\label{myProposition01}
  For random variables $X$, $Y$, and samples $x$, $y$, the following hold:
  \begin{itemize}
    \item $\left| {{r^\# }\left( {X,Y} \right)} \right| \leqslant 1$ and the equality holds if and only if $X$ and $Y$ are monotone dependent.
    \item $\left| {{r^\# }\left( {x,y} \right)} \right| \leqslant 1$ and the equality holds if and only if $x$ and $y$ are monotone dependent.
  \end{itemize}
\end{myProposition} 
An interesting question might arise in one's mind: how can the simple adjustment, replacing $\sqrt {{\mathop{\rm var}} \left( X \right){\mathop{\rm var}} \left( Y \right)} $ with $\left| {{\mathop{\rm cov}} \left( {{X^ \uparrow },{Y^ \updownarrow }} \right)} \right|$, leads to the capture range expanding from linear dependence to (nonlinear) monotone dependence? The capture range is inherited from \textit{covariance} itself. The capture range of covariance is limited neither to identical dependence as that of \textit{Concordance Correlation Coefficient}, additive dependence as that of \textit{Additivity Coefficient}, nor to linear dependence as that of Pearson's $r$. In fact, it can potentially detect and measure arbitrary monotone dependence, if scaled properly. In other words, Pearson's $r$ is also measuring nonlinear monotone dependence to some extent. The adjustment is nothing more than compensating for underestimation. 

The relationships among the above-mentioned \textit{Concordance Correlation Coefficient} (${r ^ = }$), \textit{Additivity Coefficient} (${r ^ + }$), \textit{Pearson's $r$}, and the new proposed \textit{Rearrangement Correlation} (${r ^\# }$) are depicted in Figure \ref{fig1}. 

The following proposition reveals the relationship between Pearson's $r$ and its adjusted version, i.e., Rearrangement Correlation:

\begin{myProposition}\label{myProposition02}
  For random variables $X$, $Y$, and samples $x$, $y$, the following hold:
  \begin{itemize}
    \item $\left| {{r ^\# }\left( {X,Y} \right)} \right| \geqslant \left| {r \left( {X,Y} \right)} \right|$ and the equality holds if and only if $Y\mathop  = \limits^d \alpha X + \beta$, with $\operatorname{sgn} \left( {r\left( {X,Y} \right)} \right) = \operatorname{sgn} \left( \alpha  \right)$.
    \item $\left| {{r^\# }\left( {x,y} \right)} \right| \geqslant \left| {r\left( {x,y} \right)} \right|$ and the equality holds if and only if $y$ is arbitrary permutation of $a{x} + b$, with $\operatorname{sgn} \left( {r\left( {x,y} \right)} \right) = \operatorname{sgn} \left( a \right)$.
  \end{itemize}
\end{myProposition}

Proposition \ref{myProposition02} shows that ${r ^\# }\left( {X,Y} \right)$ will revert to ${r}\left( {X,Y} \right)$ if and only if $Y\mathop  = \limits^d \alpha X + \beta$, $\operatorname{sgn} \left( {r\left( {X,Y} \right)} \right) = \operatorname{sgn} \left( \alpha  \right)$, and ${r^\# }\left( {x,y} \right)$ to  ${r}\left( {x,y} \right)$ if and only if $y$ is arbitrary permutation of $a{x} + b$, with $\operatorname{sgn} \left( {r\left( {x,y} \right)} \right) = \operatorname{sgn} \left( a \right)$. It is clear that linear dependence is special case of these conditions. Thus, ${r ^\# }$ reverts to $r$ in linear scenarios.

Another question to be asked is, do we need a new monotone measure given that rank-based measures such as Spearman's $\rho$ can already measure monotone dependence? The answer is twofold:

On the one hand, ${r^\# }$ has a higher \textit{resolution} and is more accurate. Without exception, all measures designed for monotone dependence are utilizing the order information. However, what we utilize here is the original information, rather than the ranks. Mapping numerical values to their ranks does of course produce a certain loss of information. A small difference between two values may no longer be distinguished from a large difference. With sample size $n$, there are totally $\frac{{{n^3} - n}}{6}$ possible $\rho$ values between $-1$ and $+1$, whatever raw values are and however correlated patterns differ. The \textit{resolution} of Spearman's $\rho$ might be inadequate. To take a simple example, let $x = \left( {4,3,2,1} \right)$ and 
\begin{itemize}
  \item ${y_1} = \left( {5,4,3,2.00} \right)$
  \item ${y_2} = \left( {5,4,3,3.25} \right)$
  \item ${y_3} = \left( {5,4,3,3.50} \right)$
  \item ${y_4} = \left( {5,4,3,3.75} \right)$
  \item ${y_5} = \left( {5,4,3,4.50} \right)$
\end{itemize}

Obviously, ${y_1}$ and $x$ behaves exactly in the same way, with their values getting small and small step by step. The behavior of ${y_2}$, ${y_3}$, ${y_4}$, and ${y_5}$ are becoming more and more different from that of $x$. However, the $\rho$ values are all the same for ${y_2}$, ${y_3}$ and ${y_4}$. In contrast, the ${r^\#}$ values can reveal all these differences exactly.
\begin{itemize}
  \item ${r^\# }\left( {x,{y_1}} \right) = 1.00$, $\rho \left( {x,{y_1}} \right) = 1.00$
  \item ${r^\# }\left( {x,{y_2}} \right) = 0.93$, $\rho \left( {x,{y_2}} \right) = 0.80$
  \item ${r^\# }\left( {x,{y_3}} \right) = 0.85$, $\rho \left( {x,{y_3}} \right) = 0.80$
  \item ${r^\# }\left( {x,{y_4}} \right) = 0.76$, $\rho \left( {x,{y_4}} \right) = 0.80$
  \item ${r^\# }\left( {x,{y_5}} \right) = 0.38$, $\rho \left( {x,{y_5}} \right) = 0.40$
\end{itemize}

On the other hand, ${r^\# }$ is comparable with Pearson's $r$, while the latter is not. For nonlinear monotone dependence, the value of Spearman's $\rho$ might be remarkably greater than the value of Pearson's $r$. One may attempt to search for nonlinear relationships in data by checking whether the value of $\rho$ far exceeds that of $r$. However, it might be meaningless and even impossible to compare their values directly. In cases, $\rho$ can be either greater or less than $r$, and their sign can even be different. Thus the difference $\left| \rho  \right| - \left| r \right|$ is confusing. On the contrary, the signs of ${r^\# }$ and $r$ are always the same, and $\left| {r^\# } \right|$ is always greater than or equal to $\left| r \right|$. $\left| {r^\# } \right| - \left| r \right|$ equals to 0 if and only if $y$ is arbitrary permutation of $ax + b$. Its value increases with the degree of nonlinearity.

However, Spearman's $\rho$ can also be superior to ${r^\#}$ in the sense that the former is robust to outliers while the latter is not. Rearrangement correlation is a scaled covariance, and the limitation of being non-robust to outliers is inherited from covariance itself. In fact, concordance correlation coefficient, additivity coefficient, and Pearson's $r$ are also scaled covariance measures, and none of them are robust to outliers.

To be more robust, we can also transform the raw data into their ranks before calculating ${r^\#}$. Interestingly, ${r^\#}$ becomes equivalent to Spearman's $\rho$ when calculated on ranks. Let $P$ and $Q$ be the ranks of $x$ and $y$ respectively. Then,  in the sense that $sd\left( {P,Q} \right) = sd\left( {{P^ \uparrow },{Q^ \updownarrow }} \right) = \frac{{n\left( {n + 1} \right)}}{{12}}$, ${r^\#}\left( {P,Q} \right) = \rho \left( {P,Q} \right)$. This explains why $\rho$ can measure nonlinear monotone relationships while $r$ only measures linear ones, despite them sharing a similar formula. The key is not just the ranking but achieving a sharp bound. Since $\rho$ and ${r^\#}$ are equivalent when applied to ranks, and ${r^\#}$ can measure arbitrary monotone dependence (as proven in our manuscript), $\rho$ can do the same. 

\section{Experiments}
\subsection{Performance metrics}

The main purpose of proposing Rearrangement Correlation is to provide a measure of dependence strength for nonlinear monotone relationships, rather than to simply serve as a test statistic for testing independence. Thus, our performance metrics focus on strength measurement.
 
The basic question to be asked when measuring any attribute is how accurate is this measurement, and there should be no exception for dependence measurement. ISO 5725 uses two terms \textit{``trueness''} and \textit{``precision''} to describe the accuracy of a measurement method. \textit{Trueness} refers to the closeness of agreement between the mean or median of measured results and the true or accepted reference value. \textit{Precision} refers to the closeness of agreement between measured values \citep{isoAccuracyTruenessPrecision1994}. The comprehensive performance of \textit{trueness} and \textit{precision} can be represented as the mean absolute error (\textit{MAE} for short), and calculated as the mean of the absolute values of the difference between the measured value and the conventional true value. On the whole, a measure with lower MAE value is better.

We evaluate the performance of different measures in a supervised way. We employ the coefficient of determination, ${R^2}$, which is defined as the proportion of variance for one variable explained by the other variable, as the ground truth of strength of dependence, which is common in practices \citep{reshefDetectingNovelAssociations2011}. Further, we take its square root, $R$, as the conventional true value of the relationship strength. A simple evidence is that Pearson's $r$, as the golden standard for measuring linear dependence, is equivalent to $R$, as long as the relationship is linear. Thus, it is reasonable to adopt $R$ as the reference value.

\subsection{Simulation procedure}
We investigate the accuracy performance of ${r^\# }$, along with ${r^ + }$, Pearson's $r$, Spearman's $\rho$, Kendall's $\tau$ and four other leading dependence measures, i.e., HSIC, dCor, MIC and Chatterjee's $\xi $ in the following way: for each scenario $y = f\left( x \right)$, we generated 512 pairs of $\left( {x,y} \right)$ from the regression model $y = f\left( x \right) + \varepsilon $, and computed the values of different measures between $x$ and $y$ at different $R$ levels. In the regression model, the $x$ sample is uniformly distributed on the unit interval $\left( {0,1} \right)$, and the noise is normally distributed as $\varepsilon  \sim N\left( {0,\sigma } \right)$, with $\sigma $ controlling $R$ to a certain level.
\begin{equation*}
  R = \sqrt {1 - \frac{{{\sigma ^2}}}{{{\text{var}}\left( Y \right)}}}
\end{equation*}
For the sake of robustness, the computation process is repeated 10 times for each measure at each $R$ level, and the mean value is adopted.

Simulation procedure is implemented in the \textit{recor} R package, which is available in supplemental materials. The workflow is to call \textit{accuracy\_db}() firstly, \textit{accuracy\_results\_frm\_db}() secondly and \textit{accuracy\_plot\_lite}() finally. To reproduce the results, just keep all the parameters as default. More details are available in the package help files.

\subsection{Performance in simulated scenarios}

\begin{figure*}[!tbp]
  \centering
  \includegraphics[width=0.8\linewidth]{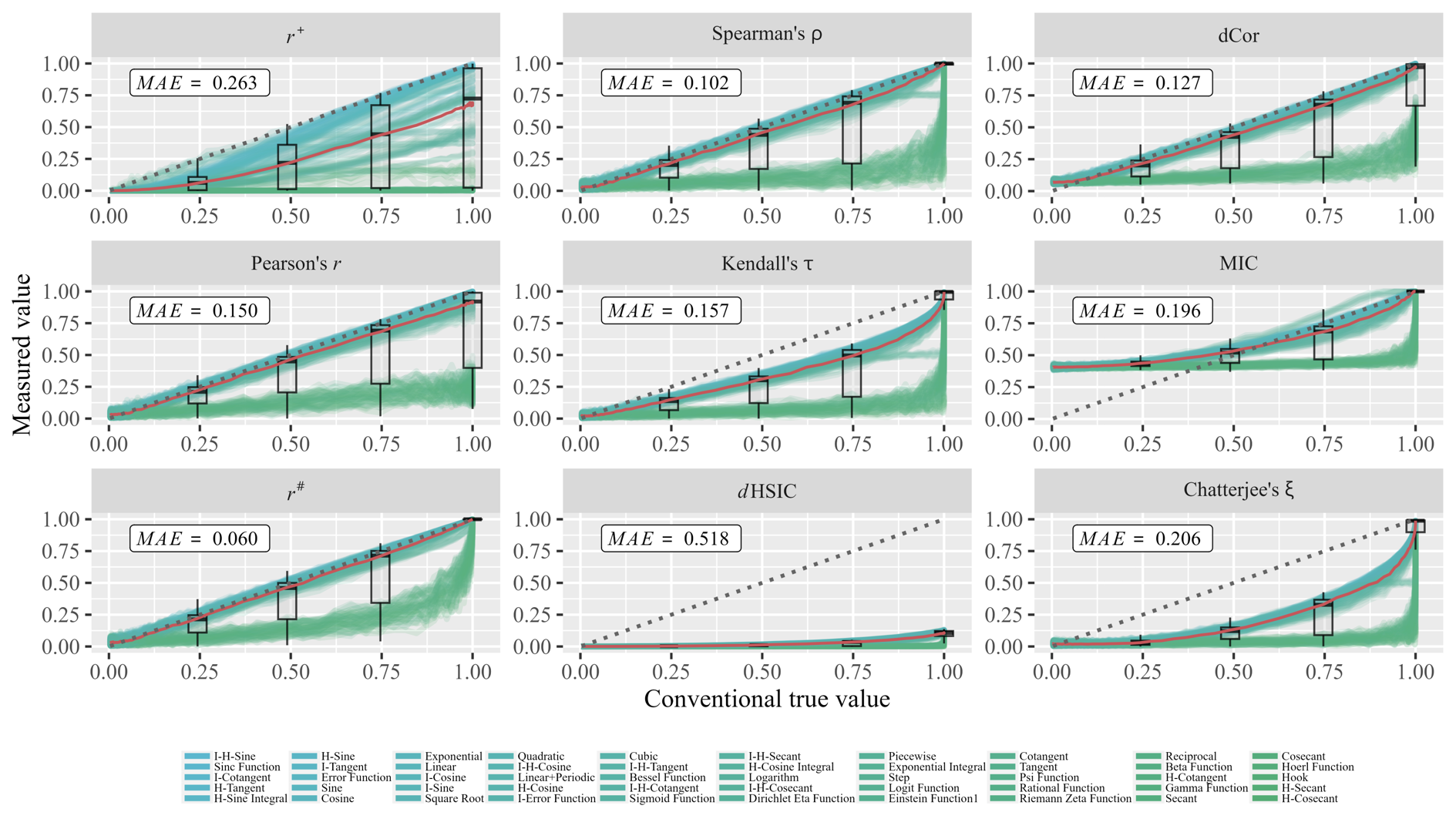}
  \caption{Performance of different measures in 50 simulated scenarios}
  \label{fig2}
\end{figure*}

The simulation is conducted in up to 50 types of different monotone scenarios, including all basic elementary functions, lots of composite functions and several special functions.  To our knowledge, our research explores the most extensive and representative range of scenarios. Detailed descriptions of these scenarios can be found in Appendix \ref{appendixSimulatedScenarios}, and the results are shown in Figure \ref{fig2}.

Figure \ref{fig2} shows the scatter plots of conventional true value versus measured values in different scenarios. The nine investigated measures are located in nine panels respectively. In each panel, there are totally 50 transparent green lines, representing the measured value ($Y$-axis) with respect to conventional true value ($X$-axis) for 50 scenarios. The prefixes \textit{I-} and \textit{H-} refer to \textit{Inverse} and \textit{Hyperbolic} respectively. For example, \textit{I-H-Tangent} stands for \textit{Inverse Hyperbolic Tangent} function. In all these panels, the dashed diagonal lines represent an ideal measure, the score of which is exactly the same as the conventional true value. Apparently, the closeness to this reference line reflects the performance of a measure for a certain scenario. The median values of each measure among scenarios at different $R$ levels are also calculated and denoted by the non-transparent red line.

We first look at the extreme values on both sides. It is expected that a measure will score nearly zero when $X$ and $Y$ are randomly generated, i.e.,  $R \approx 0$, and score one when there is perfect monotone relationship, i.e., $R = 1$. Only Spearman's $\rho$, Kendall's $\tau$ and the adjusted ${r^\# }$  meet this requirement. MIC also scores $ + 1$ when $R = 1$, however, it tends to overestimate the strength when $R$ is near zero, as also reported in other literature before \citep{chatterjeeNewCoefficientCorrelation2021}. The remaining five measures, i.e., ${r^ + }$, $r$, HSIC, dCor and $\xi $, underestimate the strength of nonlinear relationships, and never converge to $ + 1$  even when $R$ approaches to 1.

Now let's take a close look at the intermediate values. It can be seen in Figure \ref{fig2} that the non-transparent red line of  ${r^\# }$ is the closest one to the dashed line, which means the measured values by ${r^\# }$ possess the minimum error. To further quantify the accuracy, we add four boxplots at four representative $R$ levels (approximately, 0.25, 0.50, 0.75, and 1.00) for each measure. ${r^\# }$ has the highest trueness in all these representative levels. As for precision, the ${r^\# }$ also outperforms all other measures except HSIC and MIC. Although HSIC and MIC possesses the best precision, they suffer from lower trueness. HSIC tends to underestimate the strength severely, and MIC is also a biased measure, tending to overestimate the strength when the signal is weak, and underestimate it when the signal is strong, as shown in Figure \ref{fig2}. 

The overall performance in terms of MAE is ordered as $\;{r^\# }\left( {0.060} \right) \succ \rho \left( {0.102} \right) \succ dCor\left( {0.127} \right) \succ r\left( {0.150} \right) \succ \tau \left( {0.157} \right) \succ MIC\left( {0.196} \right) \succ \xi \left( {0.206} \right) \succ {r^ + }\left( {0.263} \right) \succ HSIC\left( {0.518} \right)$. ${r^\# }$ possesses significant accuracy advantage over all other measures.

\subsection{Performance in real-life scenarios}
In addition to simulated scenarios, we also investigate the performance of these measures on real life scenarios provided by NIST \citep{nationalinstituteofstandardsandtechnologyNISTStandardReference2003}. There are five available monotone scenarios: Chwirut1, Hahn1, Rat43, Roszman1, and Thurber. Details for these scenarios are available in Appendix \ref{appendixReallifeScenarios}. 

\begin{figure*}[!tbp]
  \centering
  \includegraphics[width=0.8\linewidth]{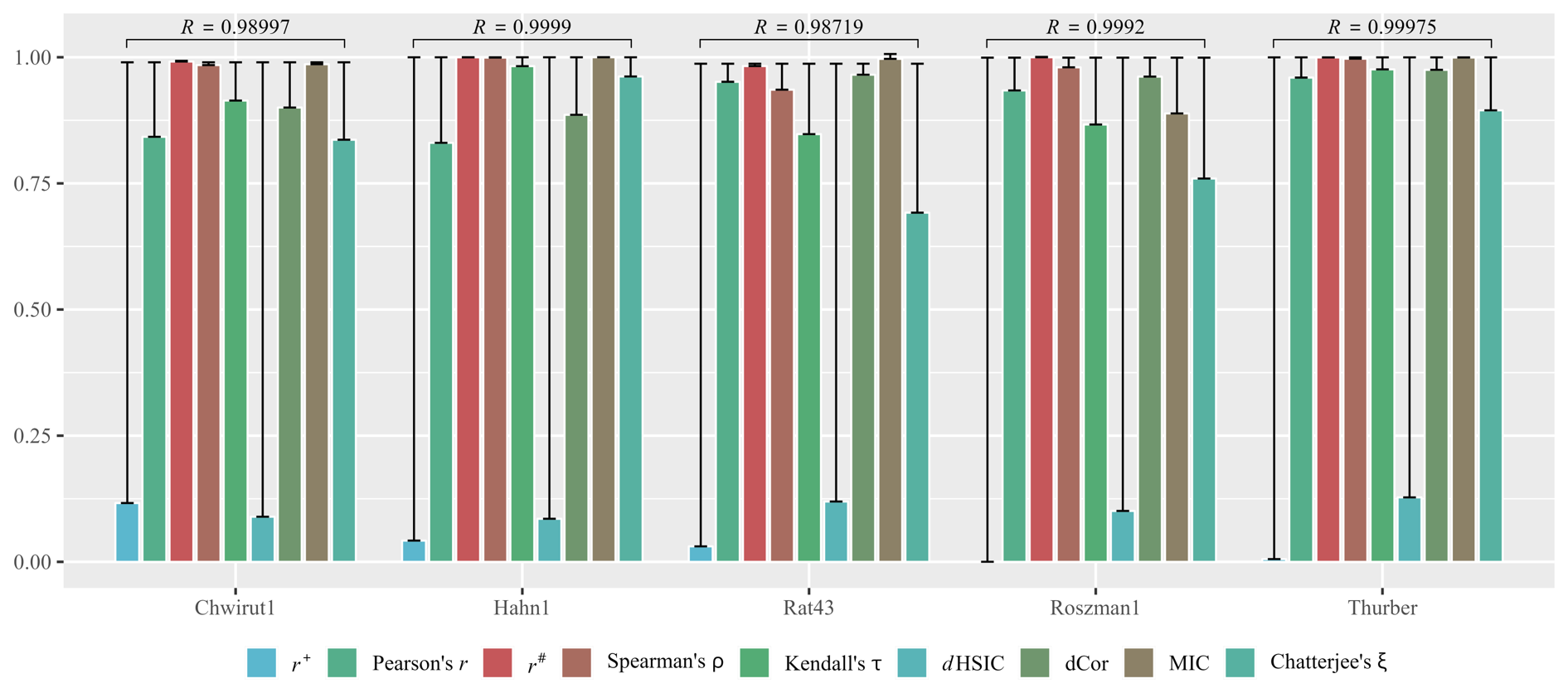}
  \caption{Performance of Different Measures in 5 Real-life Scenarios}
  \label{fig3}
\end{figure*}

The performance of nine measures in these five scenarios are depicted in Figure \ref{fig3}. Bar plots illustrate the measured values, the conventional true value verified by NIST is annotated on the top of each scenario group. And the differences between the measured value and the true value are mapped as error bars. 

It can be seen from Figure \ref{fig3} that ${r^\# }$ possesses minimum error, or best accuracy performance among all these five scenarios, with its MAE value as 0.00141, followed by $\rho $(0.0159), MIC(0.0249), dCor(0.0575),  $\tau $(0.0779), $r$(0.0916),  $\xi $(0.166), HSIC(0.891) and ${r^ + }$(0.956).

\subsection{Performance in non-monotone scenarios}
It's worth noting that the aforementioned scenarios only cover monotone cases. To evaluate performance in typical non-monotone contexts, we conducted experiments in 16 scenarios, encompassing those outlined in \citep{reshefDetectingNovelAssociations2011} and \citep{simonCommentDetectingNovel2014}. Details for these scenarios are available in Appendix \ref{appendixNonmonotoneScenarios}.

As anticipated, the MAE value of ${r^\# }$ reaches a significant 0.418, notably inferior to those of $\xi $(0.141), MIC (0.157) and dCor (0.364). In essence, ${r^\# }$ struggles to accurately measure non-monotone dependence. This limitation stems from its reliance on \textit{covariance}, which inherently fails to detect non-monotone relationships. To illustrate, consider a standard introductory text book example, i.e., $\operatorname{cov} \left( {X,Y} \right) = 0$ despite $Y$ being totally dependent on $X$ via $Y = {X^2}$. Attempts to tighten its bound proves futile. 

However, the performance of ${r^\# }$ is also superior to those of Spearman's $\rho$(0.431), Pearson's $r$(0.437) and Kendall's $\tau$(0.461). As for accuracy performance, ${r^\# }$ outperforms the three classical correlation coefficients in not only monotone, but also non-monotone scenarios.

\section{Conclusion and discussion}
We proposed here an adjusted version of Pearson's $r$, i.e., \textit{rearrangement correlation}, which can be treated as counterpart of Pearson's $r$ for
nonlinear monotone dependence. 

The basic idea of rearrangement correlation is simple and straightforward. Its mathematical foundation is a sharpened version of the famous Cauchy-Schwarz Inequality. Tighter bound leads to wider capture range. With the adjustment, the capture range of Pearson's $r$ is extended from linear dependence to (nonlinear) monotone dependence. Simulated and real-life investigations demonstrate that the rearrangement correlation is more accurate in measuring nonlinear monotone dependence than the three classical correlation coefficients and other more recently proposed dependence measures.

We may draw the conclusion that: Pearson's $r$ is undoubtedly the gold measure for linear dependence. Now, \textit{it might be the gold measure also for nonlinear monotone dependence, if adjusted.}

\printbibliography

\newpage

\appendix

\section{Appendix}
\subsection{Proofs of theorems, corollaries and propositions}\label{appendixProof}

\setcounter{myTheorem}{0}
\setcounter{myCorollary}{0}
\setcounter{myProposition}{0}

\begin{myTheorem}
  For random variables $X$ and $Y$, we have
  \begin{equation*}
    \left| {EXY} \right| \leqslant \left| {E{X^ \uparrow }{Y^ \updownarrow }} \right| \leqslant \sqrt {E{X^2}E{Y^2}}. 
  \end{equation*}
  The equality on the left holds if and only if $X$ and $Y$ are monotone dependent, and the equality on the right holds if and only if $Y\mathop  = \limits^d \alpha X$, with $\operatorname{sgn} \left( {EXY} \right) = \operatorname{sgn} \left( \alpha  \right)$.
  
  Here, $\mathop  = \limits^d $ denotes equality in distribution, and $E{X^ \uparrow }{Y^ \updownarrow }$ is defined as:
  \begin{equation*}
    E{X^ \uparrow }{Y^ \updownarrow } = \left\{ {\begin{array}{*{20}{c}}
      {E{X^ \uparrow }{Y^ \uparrow },if\;\;EXY\geqslant0} \\ 
      {E{X^ \uparrow }{Y^ \downarrow },if\;\;EXY < 0} 
    \end{array}} \right.    
  \end{equation*}
\end{myTheorem}
\begin{proof}
  The proof will be completed in two parts.
  \begin{itemize}
    \item The proof of $\left| {EXY} \right| \leqslant \left| {E{X^ \uparrow }{Y^ \updownarrow }} \right|$ is mainly based on the rearrangement theorem for functions, i.e., Theorem 378 on page 278 of \citep{hardyInequalities1952}: Let ${f^ \uparrow }$, ${g^ \uparrow }$ denote increasing rearrangements and ${f^ \downarrow }$, ${g^ \downarrow }$ decreasing rearrangements of $f$ and $g$ on $\left[ {0,1} \right]$ as defined on page 276 of \citep{hardyInequalities1952}. Then we have 
    \begin{equation*}
      \int_0^1 {{f^ \uparrow }\left( u \right){g^ \downarrow }\left( u \right)} du\leqslant\int_0^1 {f\left( u \right)g\left( u \right)} du\leqslant\int_0^1 {{f^ \uparrow }\left( u \right){g^ \uparrow }\left( u \right)} du.
    \end{equation*}
    Let $\prod \left( {F,G} \right)$ be the set of all joint cdf's on ${\mathbb{R}^2}$ having $F$ and $G$ as marginal cdf's. For arbitrary cdf $H \in \prod $ there exists $\left( {X,Y} \right)$: $\left[ {0,1} \right] \to {\mathbb{R}^2}$ such that $\left[ {X\left( U \right),Y\left( U \right)} \right]$ has cdf $H$. We can let $f\left( u \right) = X\left( u \right)$ and $g\left( u \right) = Y\left( u \right)$ so that $EXY = \int_0^1 {f\left( u \right)g\left( u \right)} du$. The increasing and decreasing rearrangements of $f$ and $g$ are just ${f^ \uparrow }\left( u \right) = {F^{ - 1}}\left( u \right)$, ${f^ \downarrow }\left( u \right) = {F^{ - 1}}\left( {1 - u} \right)$, ${g^ \uparrow }\left( u \right) = {G^{ - 1}}\left( u \right)$, and ${g^ \downarrow }\left( u \right) = {G^{ - 1}}\left( {1 - u} \right)$, as stated in \citep{whittBivariateDistributionsGiven1976}. Thus, we have
    \begin{equation*}
      E{X^ \uparrow }{Y^ \downarrow } \leqslant EXY \leqslant E{X^ \uparrow }{Y^ \uparrow }.
    \end{equation*}
    The right-hand (\textit{resp.} left-hand) equality holds if and only if $\left( {X,Y} \right)\mathop  = \limits^d \left( {{F^{ - 1}}\left( U \right),{G^{ - 1}}\left( U \right)} \right)$ (\textit{resp.} $\left( {X,Y} \right)\mathop  = \limits^d \left( {{F^{ - 1}}\left( U \right),{G^{ - 1}}\left( {1 - U} \right)} \right)$). The equality conditions can be equivalently expressed as $X$ and $Y$ are increasing (\textit{resp.} decreasing) monotone dependent \citep{mikusinskiFrechetBoundsRevisted1991}. 
    
    If $EXY \geqslant 0$, we have $E{X^ \uparrow }{Y^ \updownarrow } = E{X^ \uparrow }{Y^ \uparrow } \geqslant EXY \geqslant 0$, which implies that $\left| {E{X^ \uparrow }{Y^ \updownarrow }} \right| = E{X^ \uparrow }{Y^ \uparrow } \geqslant EXY = \left| {EXY} \right|$, and the equality holds if and only if $X$ and $Y$ are increasing monotone dependent. If $EXY < 0$, we have $E{X^ \uparrow }{Y^ \updownarrow } = E{X^ \uparrow }{Y^ \downarrow } \leqslant EXY < 0$, which implies that $\left| {E{X^ \uparrow }{Y^ \updownarrow }} \right| =  - E{X^ \uparrow }{Y^ \downarrow } \geqslant  - EXY = \left| {EXY} \right|$, and the equality holds if and only if $X$ and $Y$ are decreasing monotone dependent. Either way, we have $\left| {E{X^ \uparrow }{Y^ \updownarrow }} \right| \geqslant \left| {EXY} \right|$, and the equality holds if and only if $X$ and $Y$ are monotone dependent.
    \item The proof of $\left| {E{X^ \uparrow }{Y^ \updownarrow }} \right| \leq \sqrt {E{X^2}E{Y^2}} $ is mainly based on Cauchy-Schwarz Inequality:  If $EXY \geqslant 0$, $\left| {E{X^ \uparrow }{Y^ \updownarrow }} \right| = \left| {E{X^ \uparrow }{Y^ \uparrow }} \right|\leqslant\sqrt {E{{\left( {{X^ \uparrow }} \right)}^2}E{{\left( {{Y^ \uparrow }} \right)}^2}}  = \sqrt {E{X^2}E{Y^2}} $ in the sense that ${X^ \uparrow }\mathop  = \limits^d X$, and ${Y^ \uparrow }\mathop  = \limits^d Y$. And the equality holds if and only if ${Y^ \uparrow } = \alpha{X^ \uparrow }$, equivalently, $Y\mathop  = \limits^d {\alpha}X$, for some constant $\alpha  \geq 0$. Similarly, If $E\left( {XY} \right) < 0$, $\left| {E{X^ \uparrow }{Y^ \updownarrow }} \right| = \left| {E{X^ \uparrow }{Y^ \downarrow }} \right|\leqslant\sqrt {E{{\left( {{X^ \uparrow }} \right)}^2}E{{\left( {{Y^ \downarrow }} \right)}^2}}  = \sqrt {E{X^2}E{Y^2}} $ in the sense that ${X^ \uparrow }\mathop  = \limits^d X$, and ${Y^ \downarrow }\mathop  = \limits^d Y$. And the equality holds if and only if ${Y^ \downarrow } = \alpha{X^ \uparrow }$, equivalently, $Y\mathop  = \limits^d \alpha X$, for some constant $\alpha  < 0$. Either way, we have $\left| {E{X^ \uparrow }{Y^ \updownarrow }} \right| \leqslant \sqrt {E{X^2}E{Y^2}} $ and the equality holds if and only if $Y\mathop  = \limits^d \alpha X$, with $\operatorname{sgn} \left( {EXY} \right) = \operatorname{sgn} \left( \alpha  \right)$.
  \end{itemize}  
\end{proof}

\begin{myTheorem}
  For samples $x$ and $y$ we have
    \begin{equation*}
    \left| {\left\langle {x,y} \right\rangle } \right| \leqslant \left| {\left\langle {{x^ \uparrow },{y^ \updownarrow }} \right\rangle } \right| \leqslant \left\| x \right\|\left\| y \right\|.
  \end{equation*}
  The equality on the left holds if and only if $x$ and $y$ are monotone dependent, and the equality on the right holds if and only if $y$ is arbitrary permutation of $a{x}$, with $\operatorname{sgn} \left( {\left\langle {x,y} \right\rangle } \right) = \operatorname{sgn} \left( a \right)$.
  
  Here, $\left\langle {{x^ \uparrow },{y^ \updownarrow }} \right\rangle $ is defined as:
  \begin{equation*}
    \left\langle {{x^ \uparrow },{y^ \updownarrow }} \right\rangle  = \left\{ {\begin{array}{*{20}{c}}
      {\left\langle {{x^ \uparrow },{y^ \uparrow }} \right\rangle ,if\;\;\left\langle {x,y} \right\rangle  \geqslant 0} \\ 
      {\left\langle {{x^ \uparrow },{y^ \downarrow }} \right\rangle ,if\;\;\left\langle {x,y} \right\rangle  < 0} 
    \end{array}} \right.
  \end{equation*}
\end{myTheorem}
\begin{proof}
  The proof will also be completed in two parts.
  \begin{itemize}
    \item The proof of $\left| {\left\langle {x,y} \right\rangle } \right| \leqslant \left| {\left\langle {{x^ \uparrow },{y^ \updownarrow }} \right\rangle } \right|$ is mainly based on another rearrangement theorem for finite sets, i.e., Theorem 368 on page 261 of \citep{hardyInequalities1952}: 
    With ${x^ \uparrow } = \left( {{x_{\left( 1 \right)}},{x_{\left( 2 \right)}}, \cdots ,{x_{\left( n \right)}}} \right)$, ${y^ \uparrow } = \left( {{y_{\left( 1 \right)}},{y_{\left( 2 \right)}}, \cdots ,{y_{\left( n \right)}}} \right)$, and ${y^ \downarrow } = \left( {{y_{\left( n \right)}},{y_{\left( {n - 1} \right)}}, \cdots ,{y_{\left( 1 \right)}}} \right)$, we have
    \begin{equation*}
      \sum\limits_{i = 1}^n {{x_{\left( i \right)}}{y_{\left( {n - i + 1} \right)}}}  \leqslant \sum\limits_{i = 1}^n {{x_i}{y_i}}  \leqslant \sum\limits_{i = 1}^n {{x_{\left( i \right)}}{y_{\left( i \right)}}}.
    \end{equation*}
    That is,
    \begin{equation*}
      \left\langle {{x^ \uparrow },{y^ \downarrow }} \right\rangle  \leqslant \left\langle {x,y} \right\rangle  \leqslant \left\langle {{x^ \uparrow },{y^ \uparrow }} \right\rangle,      
    \end{equation*}
    and the right-hand (\textit{resp.} left-hand) equality holds if and only if $x$ and $y$ are similarly(\textit{resp.} oppositely) ordered. According to the definitions of ``\textit{similarly(\textit{resp.} oppositely) ordered}'' on page 43 in \citep{hardyInequalities1952}, the equality conditions can be equivalently expressed as \textit{$x$ and $y$ are increasing(\textit{resp.} decreasing) monotone dependent}.  

    If $\left\langle {x,y} \right\rangle  \geqslant 0$, we have $\left\langle {{x^ \uparrow },{y^ \updownarrow }} \right\rangle  = \left\langle {{x^ \uparrow },{y^ \uparrow }} \right\rangle  \geqslant \left\langle {x,y} \right\rangle  \geqslant 0$, which implies $\left| {\left\langle {{x^ \uparrow },{y^ \updownarrow }} \right\rangle } \right| = \left\langle {{x^ \uparrow },{y^ \uparrow }} \right\rangle  \geqslant \left\langle {x,y} \right\rangle  = \left| {\left\langle {x,y} \right\rangle } \right|$, and the equality holds if and only if $x$ and $y$ are increasing monotone dependent.If $\left\langle {x,y} \right\rangle  < 0$, we have $\left\langle {{x^ \uparrow },{y^ \updownarrow }} \right\rangle  = \left\langle {{x^ \uparrow },{y^ \downarrow }} \right\rangle  \leqslant \left\langle {x,y} \right\rangle  < 0$, which implies $\left| {\left\langle {{x^ \uparrow },{y^ \updownarrow }} \right\rangle } \right| =  - \left\langle {{x^ \uparrow },{y^ \downarrow }} \right\rangle  \geqslant  - \left\langle {x,y} \right\rangle  = \left| {\left\langle {x,y} \right\rangle } \right|$, and the equality holds if and only if $x$ and $y$ are decreasing monotone dependent. Either way, we have $\left| {\left\langle {{x^ \uparrow },{y^ \updownarrow }} \right\rangle } \right| \geqslant \left| {\left\langle {x,y} \right\rangle } \right|$ and the equality holds if and only if  $x$ and $y$ are monotone dependent.

    \item The proof of $\left| {\left\langle {{x^ \uparrow },{y^ \updownarrow }} \right\rangle } \right| \leqslant \left\| x \right\|\left\| y \right\|$ is mainly based on Cauchy-Schwarz Inequality:  in the sense that norm $\left\|  \cdot  \right\|$ is permutation invariant, we have $\left\| {{x^ \uparrow }} \right\| = \left\| x \right\|$ and $\left\| {{y^ \uparrow }} \right\| = \left\| {{y^ \downarrow }} \right\| = \left\| y \right\|$. If $\left\langle {x,y} \right\rangle  \geqslant 0$, we have $\left| {\left\langle {{x^ \uparrow },{y^ \updownarrow }} \right\rangle } \right| = \left| {\left\langle {{x^ \uparrow },{y^ \uparrow }} \right\rangle } \right| \leqslant \left\| {{x^ \uparrow }} \right\|\left\| {{y^ \uparrow }} \right\| = \left\| x \right\|\left\| y \right\|$, and the equality holds if and only if ${y^ \uparrow } = a{x^ \uparrow }$, or equivalently, $y$ is arbitrary permutation of $a{x}$ for some constant $a \geq 0$. If $\left\langle {x,y} \right\rangle  < 0$, we have $\left| {\left\langle {{x^ \uparrow },{y^ \updownarrow }} \right\rangle } \right| = \left| {\left\langle {{x^ \uparrow },{y^ \downarrow }} \right\rangle } \right|\leqslant\left\| {{x^ \uparrow }} \right\|\left\| {{y^ \downarrow }} \right\| = \left\| x \right\|\left\| y \right\|$, and the equality holds if and only if ${y^ \downarrow } = a{x^ \uparrow }$, or equivalently, $y$ is arbitrary permutation of $a{x}$ for some constant $a < 0$.  Either way, we have $\left| {\left\langle {{x^ \uparrow },{y^ \updownarrow }} \right\rangle } \right|\leqslant\left\| x \right\|\left\| y \right\|$, and the equality holds if and only if $y$ is arbitrary permutation of $a{x}$, , with $\operatorname{sgn} \left( {\left\langle {x,y} \right\rangle } \right) = \operatorname{sgn} \left( a \right)$.
  \end{itemize}
\end{proof}

\begin{myCorollary}
  For random variables $X$ and $Y$, we have covariance equality series as:
  \begin{equation*}
    \begin{aligned}
      \left| {\operatorname{cov} \left( {X,Y} \right)} \right| \leqslant \left| {\operatorname{cov} \left( {{X^ \uparrow },{Y^ \updownarrow }} \right)} \right| &\leqslant \sqrt {\operatorname{var} \left( X \right)\operatorname{var} \left( Y \right)}  \\
      &\leqslant \tfrac{1}{2}\left( {\operatorname{var} \left( X \right) + \operatorname{var} \left( Y \right)} \right) \\
      &\leqslant \tfrac{1}{2}\left( {\operatorname{var} \left( X \right) + \operatorname{var} \left( Y \right) + {{\left| {{\bar X} - {\bar Y}} \right|}^2}} \right)      
    \end{aligned}
  \end{equation*}
  The first equality holds if and only if $X$ and $Y$ are monotone dependent, and the second equality holds if and only if $Y\mathop  = \limits^d \alpha X + \beta $, with $\operatorname{sgn} \left( {\operatorname{cov} \left( {X,Y} \right)} \right) = \operatorname{sgn} \left( \alpha  \right)$.

  Here, $\operatorname{cov} \left( {{X^ \uparrow },{Y^ \updownarrow }} \right)$ is defined as:
  \begin{equation*}
    {\text{cov}}\left( {{X^ \uparrow },{Y^ \updownarrow }} \right): = \left\{ {\begin{array}{*{20}{c}}
      {{\text{cov}}\left( {{X^ \uparrow },{Y^ \uparrow }} \right),\;\;if\;\;{\text{cov}}\left( {X,Y} \right) \geqslant 0} \\ 
      {{\text{cov}}\left( {{X^ \uparrow },{Y^ \downarrow }} \right)\;\;if\;\;{\text{cov}}\left( {X,Y} \right) < 0} 
    \end{array}} \right.
  \end{equation*}
\end{myCorollary}
\begin{proof}
  $\left| {\operatorname{cov} \left( {X,Y} \right)} \right|\leqslant\left| {\operatorname{cov} \left( {{X^ \uparrow },{Y^ \updownarrow }} \right)} \right|\leqslant\sqrt {\operatorname{var} \left( X \right)\operatorname{var} \left( Y \right)} $, and the equality conditions are immediate from Theorem \ref{myTheorem01}. The remaining parts of the inequality series are obvious.  
\end{proof}

\begin{myCorollary}
  For samples $x$ and $y$, we have covariance inequality series as
  \begin{equation*}
    \begin{aligned}
    \left| {{s_{x,y}}} \right| \leqslant \left| {{s_{{x^ \uparrow },{y^ \updownarrow }}}} \right| &\leqslant \sqrt {s_x^2s_y^2}  \\
    &\leqslant \frac{1}{2}\left( {s_x^2 + s_y^2} \right) \\
    &\leqslant \frac{1}{2}\left( {s_x^2 + s_y^2 + {{\left| {\bar x - \bar y} \right|}^2}} \right)
  \end{aligned}
  \end{equation*}
  The first equality holds if and only if $x$ and $y$ are monotone dependent, and the second equality holds if and only if $y$ is arbitrary permutation of $a{x} + b$, with $\operatorname{sgn} \left( {{s_{x,y}}} \right) = \operatorname{sgn} \left( a \right)$. 

  Here, ${s_{{x^ \uparrow },{y^ \updownarrow }}}$ is defined as:
  \begin{equation*}
    {s_{{x^ \uparrow },{y^ \updownarrow }}} = \left\{ {\begin{array}{*{20}{c}}
      {{s_{{x^ \uparrow },{y^ \uparrow }}},if\;\;{s_{x,y}} \geqslant 0} \\ 
      {{s_{{x^ \uparrow },{y^ \downarrow }}},if\;\;{s_{x,y}} < 0} 
    \end{array}} \right.
  \end{equation*}
\end{myCorollary}

\begin{proof}
  $\left| {{s_{x,y}}} \right|\leqslant\left| {{s_{{x^ \uparrow },{y^ \updownarrow }}}} \right|\leqslant\sqrt {s_x^2s_y^2} $, and the equality conditions are immediate from Theorem \ref{myTheorem02}. The remaining parts of the inequality series are obvious.   
\end{proof}

\begin{myProposition}
  For random variables $X$, $Y$, and samples $x$, $y$, the following hold:
  \begin{itemize}
    \item $\left| {{r^\# }\left( {X,Y} \right)} \right| \leqslant 1$ and the equality holds if and only if $X$ and $Y$ are monotone dependent.
    \item $\left| {{r^\# }\left( {x,y} \right)} \right| \leqslant 1$ and the equality holds if and only if $x$ and $y$ are monotone dependent.
  \end{itemize}
\end{myProposition} 
\begin{proof}
  The proposition is immediate from Corollary \ref{myCorollary01} and Corollary \ref{myCorollary02}.
\end{proof}

\begin{myProposition}
  For random variables $X$, $Y$, and samples $x$, $y$, the following hold:
  \begin{itemize}
    \item $\left| {{r ^\# }\left( {X,Y} \right)} \right| \geqslant \left| {r \left( {X,Y} \right)} \right|$ and the equality holds if and only if $Y\mathop  = \limits^d \alpha X + \beta$, with $\operatorname{sgn} \left( {r\left( {X,Y} \right)} \right) = \operatorname{sgn} \left( \alpha  \right)$.
    \item $\left| {{r^\# }\left( {x,y} \right)} \right| \geqslant \left| {r\left( {x,y} \right)} \right|$ and the equality holds if and only if $y$ is arbitrary permutation of $a{x} + b$, with $\operatorname{sgn} \left( {r\left( {x,y} \right)} \right) = \operatorname{sgn} \left( a \right)$.
  \end{itemize}
\end{myProposition}
\begin{proof}
  The proposition is immediate from Corollary \ref{myCorollary01} and Corollary \ref{myCorollary02}.
\end{proof}

\subsection{Experiments settings}\label{appendixExperimentsSettings}
All the experiments are implemented with the R language \citep{rcoreteamLanguageEnvironmentStatistical2024}, along with several add-on packages. The following are lists of packages and functions for the implementation of the nine measures involved in our study:
\begin{itemize}
    \item ${r^ + }$, recor::loose\_pearson()
    \item $r$,  stats::cor()
    \item ${r^\# }$, recor::sharp\_pearson()
    \item $\rho$, stats::cor(), with the argument \textit{method} set as ``spearman''
    \item $\tau$, stats::cor(), with the argument \textit{method} set as ``kendall''
    \item HSIC, dHSIC::dhsic()
    \item dCor, energy::dcor()
    \item MIC, minerva::mine\_stat()
    \item $\xi $, XICOR::calculateXI()
\end{itemize}

For convenience, we developed an R package \textit{recor}, which encapsulates all these measures as \textit{cor\_XXX}() functions. The \textit{recor} package is available as \textit{recor\_1.0.2.tar.gz} in supplemental materials. For a latest version, please visit https://github.com/byaxb/recor.

Hardware environment configuration for this study was: DELL OptiPlex 7070 Tower, equipped with 8-core CPU Core i7-9700 @ 3.00GHz, 24G DDR4 2666MHz RAM. Under this configuration, it took about 5 days to complete all the experiments.

\subsection{Simulated scenarios}\label{appendixSimulatedScenarios}
We carry out our experiments on 50 simulated scenarios, including all basic elementary functions, lots of composite functions and several typical special functions.
\begin{enumerate}
    \item Linear: $y = 2x + 1,x \in \left[ {0,1} \right]$
    \item Quadratic [asymmetry]: $y = {x^2},x \in \left[ {0,1} \right]$
    \item Square Root: $y = \sqrt x ,x \in \left[ {0,1} \right]$
    \item Cubic: $y = {x^3},x \in \left[ {0,1} \right]$
    \item Reciprocal: $y = \frac{1}{x},x \in \left[ {0,1} \right]$
    \item Exponential: $y = {e^x}$, with $x \in \left[ {0,1} \right]$
    \item Logarithm: $y = \ln x,x \in \left[ {0,1} \right]$
    \item Sine [quarter period]: $y = \sin \left( x \right),x \in \left[ {0,\frac{\pi }{2}} \right]$
    \item Cosine [quarter period]: $y = \cos \left( x \right),x \in \left[ {0,\frac{\pi }{2}} \right]$
    \item Tangent [half period]: $y = \tan \left( x \right),x \in \left[ {0,\frac{\pi }{2}} \right]$
    \item Cotangent [half period]: $y = \cot \left( x \right),x \in \left[ {0,\frac{\pi }{2}} \right]$
    \item Inverse Sine: $y = \arcsin \left( x \right),x \in \left[ {0,1} \right]$
    \item Inverse Cosine: $y = \arccos \left( x \right),x \in \left[ {0,1} \right]$
    \item Inverse Tangent: $y = \arctan \left( x \right),x \in \left[ {0,1} \right]$
    \item Inverse Cotangent: $y = {\rm{arccot}}\left( x \right),x \in \left[ {0,1} \right]$
    \item Secant [quarter period]: $y = \sec \left( x \right),x \in \left[ {0,\frac{\pi }{2}} \right]$
    \item Cosecant [quarter period]: $y = \csc \left( x \right),x \in \left[ {0,\frac{\pi }{2}} \right]$
    \item Hyperbolic Sine: $y = \sinh x = \frac{{{e^x} - {e^{ - x}}}}{2},x \in \left[ {0,1} \right]$
    \item Hyperbolic Cosine: $y = \cosh x = \frac{{{e^x} + {e^{ - x}}}}{2},x \in \left[ {0,1} \right]$
    \item Hyperbolic Tangent: $y = \tanh x = \frac{{{e^{2x}} - 1}}{{{e^{2x}} + 1}},x \in \left[ {0,1} \right]$
    \item Hyperbolic Cotangent: $y = \coth x = \frac{{{e^{2x}} + 1}}{{{e^{2x}} - 1}},x \in \left[ {0,1} \right]$
    \item Hyperbolic Secant: $y = {\rm{sech}}\left( x \right) = \frac{2}{{{e^x} + {e^{ - x}}}},x \in \left[ {0,100} \right]$
    \item Hyperbolic Cosecant: $y = csch\left( x \right) = \frac{2}{{{e^x} - {e^{ - x}}}},x \in \left[ {0,100} \right]$
    \item Inverse Hyperbolic Sine: $y = arcsinh\left( x \right) = \ln \left( {x + \sqrt {{x^2} + 1} } \right),x \in \left[ {0,1} \right]$
    \item Inverse Hyperbolic Cosine: ${y = arccosh\left( x \right) = \ln \left( {x + \sqrt {{x^2} - 1} } \right),x \in \left[ {1,2} \right]}$
    \item Inverse Hyperbolic Tangent: $\begin{array}{*{20}{l}}
        {y = arctanh\left( x \right) = \frac{1}{2}\ln \left( {\frac{{1 + x}}{{1 - x}}} \right)}
        \end{array},x \in \left[ {0,1} \right]$
    \item Inverse Hyperbolic Cotangent: $\begin{array}{*{20}{l}}
        {y = arccoth\left( x \right) = \frac{1}{2}\ln \left( {\frac{{x + 1}}{{x - 1}}} \right),x \in \left[ {1,2} \right]}
        \end{array}$
    \item Inverse Hyperbolic Secant: $\begin{array}{*{20}{l}}
        {y = arcsech\left( x \right) = \ln \left( {\frac{1}{x} + \sqrt {\frac{1}{{{x^2}}} - 1} } \right),x \in \left[ {0,1} \right]}
        \end{array}$
    \item Inverse Hyperbolic Cosecant: $\begin{array}{*{20}{l}}
        {y = arccsch\left( x \right) = \ln \left( {\frac{1}{x} + \sqrt {\frac{1}{{{x^2}}} + 1} } \right),x \in \left[ {0,1} \right]}
        \end{array}$
    \item Hook: $y = ax + \frac{b}{x},a = 1,b = 1,x \in \left[ {0,1} \right]$
    \item Rational: $y = \frac{{x + 1}}{{x - 1}},x \in \left[ {0,1} \right]$
    \item Hoerl: $y = {x^a}{e^x},a =  - 1,x \in \left[ {0,1} \right]$
    \item Sigmoid: $y = \frac{1}{{1 + {e^{ - x}}}},x \in \left[ { - 0.5,0.5} \right]$
    \item Logit: $y = \ln \frac{x}{{1 - x}},x \in \left[ {0,1} \right]$
    \item Step: $y = \left\{ {\begin{array}{*{20}{c}}
        {0,if\;0 \leqslant x < \frac{1}{2}}\\
        {\begin{array}{*{20}{c}}
        {1,if\;\frac{1}{2} \leqslant x \leqslant 1}
        \end{array}}
        \end{array}} \right.$
    \item Piecewise [Sigmoid]: $y = \left\{ {\begin{array}{*{20}{c}}
        {0,\;\;\;\;\;\;\;\;\;\;\;if\;0 \leqslant x \leqslant \frac{{49}}{{100}}}\\
        {50\left( {x - \frac{1}{2}} \right) + \frac{1}{2},if\;\frac{{49}}{{100}} < x < \frac{{51}}{{100}}}\\
        {1,\;\;\;\;\;\;\;\;\;\;if\;\frac{{51}}{{100}} \leqslant x \leqslant 1}
        \end{array}} \right.$
    \item Linear + Periodic, High Freq: $\begin{array}{*{20}{l}}
        {y = \frac{1}{{10}}\sin \left( {10.6\left( {2x - 1} \right)} \right) + \frac{{11}}{{10}}\left( {2x - 1} \right),x \in \left[ {0,1} \right]}
        \end{array}$
    \item Sinc Function: ${S_{k,h}}\left( x \right) = \frac{{\sin \left( {\pi \left( {x - kh} \right)/h} \right)}}{{\pi \left( {x - kh} \right)/h}},k = 0,h = 1,x \in \left[ {0,1} \right]$
    \item Einstein Function: ${\rm{Einstei}}{{\rm{n}}_1}\left( x \right) = \frac{{{x^2}{e^x}}}{{{{\left( {{e^x} - 1} \right)}^2}}},x \in \left[ {0,1} \right]$
    \item Exponential Integral: ${E_1}\left( x \right) = \int_x^\infty  {\frac{{{e^{ - t}}}}{t}dt} ,x \in \left[ {0,1} \right]$
    \item Hyperbolic Sine Integral: $Shi\left( x \right) = \int_0^x {\frac{{\sinh t}}{t}dt} ,x \in \left[ {0,1} \right]$
    \item Hyperbolic Cosine Integral: $Chi\left( x \right) = \gamma  + \ln x + \int_0^x {\frac{{\cosh t - 1}}{t}dt} ,x \in \left[ {0,1} \right]$. Here $\gamma $ is Euler's Constant
    \item Error Function: $erf\left( x \right) = \frac{2}{{\sqrt \pi  }}\int_0^x {{e^{ - {t^2}}}dt} ,x \in \left[ {0,1} \right]$
    \item Inverse Error Function: $inverf\left( x \right) = t + \frac{1}{3}{t^3} + \frac{7}{{30}}{t^5} +  \cdots ,t = \frac{1}{2}\sqrt \pi  x,x \in \left[ {0,1} \right]$
    \item Gamma Function: $\Gamma \left( x \right) = \int_0^\infty  {{t^{x - 1}}{e^{ - t}}dt} ,x \in \left[ {0,1} \right]$
    \item Psi Function: $\psi \left( x \right) = \frac{{{d^{k + 1}}}}{{d{x^{k + 1}}}}\ln \Gamma \left( x \right) = \frac{{\Gamma '\left( x \right)}}{{\Gamma \left( x \right)}},x \in \left[ {0,1} \right],k = 0$
    \item Riemann Zeta Function: $\zeta \left( x \right) = \sum\limits_{n = 1}^\infty  {\frac{1}{{{n^x}}}} ,x \in \left[ {0,1} \right]$
    \item Bessel Function: ${Y_v}\left( x \right) = \frac{{{J_v}\left( x \right)\cos \left( {v\pi } \right) - {J_{ - v}}\left( x \right)}}{{\sin \left( {v\pi } \right)}}$, ${J_v}\left( x \right) = {\left( {\frac{1}{2}x} \right)^v}\sum\limits_{k = 0}^\infty  {{{\left( { - 1} \right)}^k}\frac{{{{\left( {\frac{1}{4}{x^2}} \right)}^k}}}{{k!\Gamma \left( {v + k + 1} \right)}}}$, $v = 0$, $x \in \left[ {0,1} \right]$
    \item Beta Function: $B\left( {x,w} \right) = \frac{{\Gamma \left( x \right)\Gamma \left( w \right)}}{{\Gamma \left( {x + w} \right)}},w = 1,x \in \left[ {0,1} \right]$
    \item Dirichlet Eta Function: $\eta \left( x \right) = \sum\limits_{n = 1}^\infty  {\frac{{ - {1^{n - 1}}}}{{{n^x}}}} ,x \in \left[ {0,1} \right]$

\end{enumerate}

\subsection{Real-life scenarios}\label{appendixReallifeScenarios}
All the five real life scenarios are provided by NIST \citep{nationalinstituteofstandardsandtechnologyNISTStandardReference2003} as follows:
\begin{itemize}
  \item Chwirut1: ultrasonic calibration, with $Y$ as ultrasonic response, and $X$ as metal distance.
  \item Hahn1: thermal expansion of copper, with $Y$ as the coefficient of thermal expansion, and $X$ as temperature in degrees kelvin.
  \item Rat43: sigmoid growth, with $Y$ as dry weight of onion bulbs and tops, and $X$ as growing time.
  \item Roszman1: quantum defects in iodine atoms, with $Y$ as the number of quantum defects, and $X$ as the excited energy state.
  \item Thurber: semiconductor electron mobility, with $Y$ as a measure of electron mobility, and $X$ as the natural log of the density.
\end{itemize}
Data and details about these scenarios are available publicly at:\\
https://www.itl.nist.gov/div898/strd/nls/nls\_main.shtml
    
\subsection{Non-monotone scenarios}\label{appendixNonmonotoneScenarios}
We conducted our experiments on 16 non-monotone scenarios, comprehensively covering all the scenarios from \citep{reshefDetectingNovelAssociations2011} and \citep{simonCommentDetectingNovel2014}.
\begin{enumerate}
  \item Quadratic [symmetry]: $y = 4{x^2},x \in \left[ { - \frac{1}{2},\frac{1}{2}} \right]$
  \item Cubic 2: $y = 128{\left( {x - \frac{1}{3}} \right)^3} - 48{\left( {x - \frac{1}{3}} \right)^2} - 12\left( {x - \frac{1}{3}} \right),x \in \left[ {0,1} \right]$
  \item Sine, High Freq: $y = \sin \left( {16\pi x} \right),x \in \left[ {0,1} \right]$
  \item Cosine [High Freq]: $y = \cos \left( {14\pi x} \right),x \in \left[ {0,1} \right]$
  \item Lopsided L-shaped: $y = \left\{ {\begin{array}{*{20}{c}}
      {200x,\;if\;0 \leqslant x < \frac{1}{{200}}}\\
      { - 198x + \frac{{199}}{{100}},if\;\frac{1}{{200}} \leqslant x < \frac{1}{{100}}}\\
      {\begin{array}{*{20}{c}}
      { - \frac{x}{{99}} + \frac{1}{{99}},\;if\;\frac{1}{{100}} \leqslant x \leqslant 1}
      \end{array}}
      \end{array}} \right.$
  \item Circle: $y = \sqrt {1{\rm{ - }}{{\left( {2x - 1} \right)}^2}} ,x \in \left[ {0,1} \right]$
  \item Linear + Periodic, Medium Freq: $y = \sin \left( {10\pi x} \right) + x,x \in \left[ {0,1} \right]$
  \item Cubic 3: $y = 4{x^3} + {x^2} - 4x,x \in \left[ { - 1.1,1.3} \right]$
  \item Cubic, Y-stretched: $y = 41\left( {4{x^3} + {x^2} - 4x} \right),x \in \left[ { - 1.1,1.3} \right]$
  \item Sine [Two periods]: $y = \sin \left( {4\pi x} \right),x \in \left[ {0,1} \right]$
  \item Sine [Low Freq]: $y = \sin \left( {8\pi x} \right),x \in \left[ {0,1} \right]$
  \item Sine, Non-Fourier Freq [Low]: $y = \sin \left( {9\pi x} \right),x \in \left[ {0,1} \right]$
  \item Cosine, Non-Fourier Freq [Low]: $y = \cos \left( {7\pi x} \right),x \in \left[ {0,1} \right]$
  \item Sine, Varying Freq [Medium]: $y = \sin \left( {6\pi x\left( {1 + x} \right)} \right),x \in \left[ {0,1} \right]$
  \item Cosine, Varying Freq [Medium]: $y = \cos \left( {5\pi x\left( {1 + x} \right)} \right),x \in \left[ {0,1} \right]$
  \item Linear + Periodic, High Freq 2: $y = \frac{1}{5}\sin \left( {10.6\left( {2x - 1} \right)} \right) + \frac{{11}}{{10}}\left( {2x - 1} \right),x \in \left[ {0,1} \right]$

\end{enumerate}


\newpage
\section*{NeurIPS Paper Checklist}

\begin{enumerate}

\item {\bf Claims}
    \item[] Question: Do the main claims made in the abstract and introduction accurately reflect the paper's contributions and scope?
    \item[] Answer: \answerYes{} 
    \item[] Justification: The contributions are summarized in \textit{Abstract} and the last paragraph of \textit{Introduction}.
    \item[] Guidelines:
    \begin{itemize}
        \item The answer NA means that the abstract and introduction do not include the claims made in the paper.
        \item The abstract and/or introduction should clearly state the claims made, including the contributions made in the paper and important assumptions and limitations. A No or NA answer to this question will not be perceived well by the reviewers. 
        \item The claims made should match theoretical and experimental results, and reflect how much the results can be expected to generalize to other settings. 
        \item It is fine to include aspirational goals as motivation as long as it is clear that these goals are not attained by the paper. 
    \end{itemize}

\item {\bf Limitations}
    \item[] Question: Does the paper discuss the limitations of the work performed by the authors?
    \item[] Answer: \answerYes{} 
    \item[] Justification: The proposed ${r^\# }$ measures linear and nonlinear monotone relationships accurately. However, it may fail to measure non-monotone dependence. The limitations are discussed in \textit{3.5 Performance in non-monotone scenarios}.
    \item[] Guidelines:
    \begin{itemize}
        \item The answer NA means that the paper has no limitation while the answer No means that the paper has limitations, but those are not discussed in the paper. 
        \item The authors are encouraged to create a separate "Limitations" section in their paper.
        \item The paper should point out any strong assumptions and how robust the results are to violations of these assumptions (e.g., independence assumptions, noiseless settings, model well-specification, asymptotic approximations only holding locally). The authors should reflect on how these assumptions might be violated in practice and what the implications would be.
        \item The authors should reflect on the scope of the claims made, e.g., if the approach was only tested on a few datasets or with a few runs. In general, empirical results often depend on implicit assumptions, which should be articulated.
        \item The authors should reflect on the factors that influence the performance of the approach. For example, a facial recognition algorithm may perform poorly when image resolution is low or images are taken in low lighting. Or a speech-to-text system might not be used reliably to provide closed captions for online lectures because it fails to handle technical jargon.
        \item The authors should discuss the computational efficiency of the proposed algorithms and how they scale with dataset size.
        \item If applicable, the authors should discuss possible limitations of their approach to address problems of privacy and fairness.
        \item While the authors might fear that complete honesty about limitations might be used by reviewers as grounds for rejection, a worse outcome might be that reviewers discover limitations that aren't acknowledged in the paper. The authors should use their best judgment and recognize that individual actions in favor of transparency play an important role in developing norms that preserve the integrity of the community. Reviewers will be specifically instructed to not penalize honesty concerning limitations.
    \end{itemize}

\item {\bf Theory Assumptions and Proofs}
    \item[] Question: For each theoretical result, does the paper provide the full set of assumptions and a complete (and correct) proof?
    \item[] Answer: \answerYes{} 
    \item[] Justification: Full set of assumptions and complete and correct proofs are provided in \textit{2.3 New inequality tighter than Cauchy-Schwarz Inequality}, \textit{2.4 The proposed Rearrangement Correlation} and \textit{A.1 Proofs of theorems, corollaries and propositions}.
    \item[] Guidelines:
    \begin{itemize}
        \item The answer NA means that the paper does not include theoretical results. 
        \item All the theorems, formulas, and proofs in the paper should be numbered and cross-referenced.
        \item All assumptions should be clearly stated or referenced in the statement of any theorems.
        \item The proofs can either appear in the main paper or the supplemental material, but if they appear in the supplemental material, the authors are encouraged to provide a short proof sketch to provide intuition. 
        \item Inversely, any informal proof provided in the core of the paper should be complemented by formal proofs provided in appendix or supplemental material.
        \item Theorems and Lemmas that the proof relies upon should be properly referenced. 
    \end{itemize}

    \item {\bf Experimental Result Reproducibility}
    \item[] Question: Does the paper fully disclose all the information needed to reproduce the main experimental results of the paper to the extent that it affects the main claims and/or conclusions of the paper (regardless of whether the code and data are provided or not)?
    \item[] Answer: \answerYes{} 
    \item[] Justification: We fully disclosed all the information needed to reproduce the experimental results, as depicted in \textit{3.2 Simulation procedure}. 
    \item[] Guidelines:
    \begin{itemize}
        \item The answer NA means that the paper does not include experiments.
        \item If the paper includes experiments, a No answer to this question will not be perceived well by the reviewers: Making the paper reproducible is important, regardless of whether the code and data are provided or not.
        \item If the contribution is a dataset and/or model, the authors should describe the steps taken to make their results reproducible or verifiable. 
        \item Depending on the contribution, reproducibility can be accomplished in various ways. For example, if the contribution is a novel architecture, describing the architecture fully might suffice, or if the contribution is a specific model and empirical evaluation, it may be necessary to either make it possible for others to replicate the model with the same dataset, or provide access to the model. In general. releasing code and data is often one good way to accomplish this, but reproducibility can also be provided via detailed instructions for how to replicate the results, access to a hosted model (e.g., in the case of a large language model), releasing of a model checkpoint, or other means that are appropriate to the research performed.
        \item While NeurIPS does not require releasing code, the conference does require all submissions to provide some reasonable avenue for reproducibility, which may depend on the nature of the contribution. For example
        \begin{enumerate}
            \item If the contribution is primarily a new algorithm, the paper should make it clear how to reproduce that algorithm.
            \item If the contribution is primarily a new model architecture, the paper should describe the architecture clearly and fully.
            \item If the contribution is a new model (e.g., a large language model), then there should either be a way to access this model for reproducing the results or a way to reproduce the model (e.g., with an open-source dataset or instructions for how to construct the dataset).
            \item We recognize that reproducibility may be tricky in some cases, in which case authors are welcome to describe the particular way they provide for reproducibility. In the case of closed-source models, it may be that access to the model is limited in some way (e.g., to registered users), but it should be possible for other researchers to have some path to reproducing or verifying the results.
        \end{enumerate}
    \end{itemize}

\item {\bf Open access to data and code}
    \item[] Question: Does the paper provide open access to the data and code, with sufficient instructions to faithfully reproduce the main experimental results, as described in supplemental material?
    \item[] Answer: \answerYes{} 
    \item[] Justification: The data and code are available as an attached zip file, \textit{Code and data.zip}. We developed an R package \textit{recor}, which implemented all the experiments. The \textit{recor} package is available as \textit{recor\_1.0.2.tar.gz}, included in the zip file. Sufficient instructions to faithfully reproduce the experiments are available in the package help files.
    \item[] Guidelines:
    \begin{itemize}
        \item The answer NA means that paper does not include experiments requiring code.
        \item Please see the NeurIPS code and data submission guidelines (\url{https://nips.cc/public/guides/CodeSubmissionPolicy}) for more details.
        \item While we encourage the release of code and data, we understand that this might not be possible, so “No” is an acceptable answer. Papers cannot be rejected simply for not including code, unless this is central to the contribution (e.g., for a new open-source benchmark).
        \item The instructions should contain the exact command and environment needed to run to reproduce the results. See the NeurIPS code and data submission guidelines (\url{https://nips.cc/public/guides/CodeSubmissionPolicy}) for more details.
        \item The authors should provide instructions on data access and preparation, including how to access the raw data, preprocessed data, intermediate data, and generated data, etc.
        \item The authors should provide scripts to reproduce all experimental results for the new proposed method and baselines. If only a subset of experiments are reproducible, they should state which ones are omitted from the script and why.
        \item At submission time, to preserve anonymity, the authors should release anonymized versions (if applicable).
        \item Providing as much information as possible in supplemental material (appended to the paper) is recommended, but including URLs to data and code is permitted.
    \end{itemize}

\item {\bf Experimental Setting/Details}
    \item[] Question: Does the paper specify all the training and test details (e.g., data splits, hyperparameters, how they were chosen, type of optimizer, etc.) necessary to understand the results?
    \item[] Answer: \answerYes{} 
    \item[] Justification: The experimental settings are provided in \textit{3 Experiments}, and \textit{A.2 Experiments Settings}. 
    \item[] Guidelines:
    \begin{itemize}
        \item The answer NA means that the paper does not include experiments.
        \item The experimental setting should be presented in the core of the paper to a level of detail that is necessary to appreciate the results and make sense of them.
        \item The full details can be provided either with the code, in appendix, or as supplemental material.
    \end{itemize}

\item {\bf Experiment Statistical Significance}
    \item[] Question: Does the paper report error bars suitably and correctly defined or other appropriate information about the statistical significance of the experiments?
    \item[] Answer: \answerYes{} 
    \item[] Justification: We evaluate the performance of proposed method and others according to ISO 5725. Please see \textit{3 Experiments} for details.
    \item[] Guidelines:
    \begin{itemize}
        \item The answer NA means that the paper does not include experiments.
        \item The authors should answer "Yes" if the results are accompanied by error bars, confidence intervals, or statistical significance tests, at least for the experiments that support the main claims of the paper.
        \item The factors of variability that the error bars are capturing should be clearly stated (for example, train/test split, initialization, random drawing of some parameter, or overall run with given experimental conditions).
        \item The method for calculating the error bars should be explained (closed form formula, call to a library function, bootstrap, etc.)
        \item The assumptions made should be given (e.g., Normally distributed errors).
        \item It should be clear whether the error bar is the standard deviation or the standard error of the mean.
        \item It is OK to report 1-sigma error bars, but one should state it. The authors should preferably report a 2-sigma error bar than state that they have a 96\% CI, if the hypothesis of Normality of errors is not verified.
        \item For asymmetric distributions, the authors should be careful not to show in tables or figures symmetric error bars that would yield results that are out of range (e.g. negative error rates).
        \item If error bars are reported in tables or plots, The authors should explain in the text how they were calculated and reference the corresponding figures or tables in the text.
    \end{itemize}

\item {\bf Experiments Compute Resources}
    \item[] Question: For each experiment, does the paper provide sufficient information on the computer resources (type of compute workers, memory, time of execution) needed to reproduce the experiments?
    \item[] Answer: \answerYes{} 
    \item[] Justification: Details about the hardware configuration and the time of execution are shown in section ``\textit{A.2 Experiments Settings}''. 
    \item[] Guidelines:
    \begin{itemize}
        \item The answer NA means that the paper does not include experiments.
        \item The paper should indicate the type of compute workers CPU or GPU, internal cluster, or cloud provider, including relevant memory and storage.
        \item The paper should provide the amount of compute required for each of the individual experimental runs as well as estimate the total compute. 
        \item The paper should disclose whether the full research project required more compute than the experiments reported in the paper (e.g., preliminary or failed experiments that didn't make it into the paper). 
    \end{itemize}
    
\item {\bf Code Of Ethics}
    \item[] Question: Does the research conducted in the paper conform, in every respect, with the NeurIPS Code of Ethics \url{https://neurips.cc/public/EthicsGuidelines}?
    \item[] Answer: \answerYes{} 
    \item[] Justification: We have thoroughly reviewed the NeurIPS Code of Ethics. And we confirm that our research fully complies with all of its provisions.
    \item[] Guidelines:
    \begin{itemize}
        \item The answer NA means that the authors have not reviewed the NeurIPS Code of Ethics.
        \item If the authors answer No, they should explain the special circumstances that require a deviation from the Code of Ethics.
        \item The authors should make sure to preserve anonymity (e.g., if there is a special consideration due to laws or regulations in their jurisdiction).
    \end{itemize}

\item {\bf Broader Impacts}
    \item[] Question: Does the paper discuss both potential positive societal impacts and negative societal impacts of the work performed?
    \item[] Answer:  \answerNA{} 
    \item[] Justification: What we proposed here is an adjusted version of Pearson's $r$. This study simply provides theoretical results for measuring dependence and does not involve societal impacts. 
    \item[] Guidelines:
    \begin{itemize}
        \item The answer NA means that there is no societal impact of the work performed.
        \item If the authors answer NA or No, they should explain why their work has no societal impact or why the paper does not address societal impact.
        \item Examples of negative societal impacts include potential malicious or unintended uses (e.g., disinformation, generating fake profiles, surveillance), fairness considerations (e.g., deployment of technologies that could make decisions that unfairly impact specific groups), privacy considerations, and security considerations.
        \item The conference expects that many papers will be foundational research and not tied to particular applications, let alone deployments. However, if there is a direct path to any negative applications, the authors should point it out. For example, it is legitimate to point out that an improvement in the quality of generative models could be used to generate deepfakes for disinformation. On the other hand, it is not needed to point out that a generic algorithm for optimizing neural networks could enable people to train models that generate Deepfakes faster.
        \item The authors should consider possible harms that could arise when the technology is being used as intended and functioning correctly, harms that could arise when the technology is being used as intended but gives incorrect results, and harms following from (intentional or unintentional) misuse of the technology.
        \item If there are negative societal impacts, the authors could also discuss possible mitigation strategies (e.g., gated release of models, providing defenses in addition to attacks, mechanisms for monitoring misuse, mechanisms to monitor how a system learns from feedback over time, improving the efficiency and accessibility of ML).
    \end{itemize}
    
\item {\bf Safeguards}
    \item[] Question: Does the paper describe safeguards that have been put in place for responsible release of data or models that have a high risk for misuse (e.g., pretrained language models, image generators, or scraped datasets)?
    \item[] Answer: \answerNA{} 
    \item[] Justification: Neither the proposed method nor the released data has risk for misuse. The applicability range of this method has undergone comprehensive discussion in \textit{3.5 Performance in non-monotone scenarios}.
    \item[] Guidelines:
    \begin{itemize}
        \item The answer NA means that the paper poses no such risks.
        \item Released models that have a high risk for misuse or dual-use should be released with necessary safeguards to allow for controlled use of the model, for example by requiring that users adhere to usage guidelines or restrictions to access the model or implementing safety filters. 
        \item Datasets that have been scraped from the Internet could pose safety risks. The authors should describe how they avoided releasing unsafe images.
        \item We recognize that providing effective safeguards is challenging, and many papers do not require this, but we encourage authors to take this into account and make a best faith effort.
    \end{itemize}

\item {\bf Licenses for existing assets}
    \item[] Question: Are the creators or original owners of assets (e.g., code, data, models), used in the paper, properly credited and are the license and terms of use explicitly mentioned and properly respected?
    \item[] Answer: \answerYes{} 
    \item[] Justification: All code and data used in our paper are properly credited and explicitly mentioned, as shown in \textit{A.2 Experiments Settings}.
    \item[] Guidelines:
    \begin{itemize}
        \item The answer NA means that the paper does not use existing assets.
        \item The authors should cite the original paper that produced the code package or dataset.
        \item The authors should state which version of the asset is used and, if possible, include a URL.
        \item The name of the license (e.g., CC-BY 4.0) should be included for each asset.
        \item For scraped data from a particular source (e.g., website), the copyright and terms of service of that source should be provided.
        \item If assets are released, the license, copyright information, and terms of use in the package should be provided. For popular datasets, \url{paperswithcode.com/datasets} has curated licenses for some datasets. Their licensing guide can help determine the license of a dataset.
        \item For existing datasets that are re-packaged, both the original license and the license of the derived asset (if it has changed) should be provided.
        \item If this information is not available online, the authors are encouraged to reach out to the asset's creators.
    \end{itemize}

\item {\bf New Assets}
    \item[] Question: Are new assets introduced in the paper well documented and is the documentation provided alongside the assets?
    \item[] Answer: \answerYes{} 
    \item[] Justification: We developed an R package to reproduce the experiments. And all the functions in this package are well documented, which can be accessed by the command ``?function\_name'' after installation. For more details, see \textit{A.2 Experiments Settings}.
    \item[] Guidelines:
    \begin{itemize}
        \item The answer NA means that the paper does not release new assets.
        \item Researchers should communicate the details of the dataset/code/model as part of their submissions via structured templates. This includes details about training, license, limitations, etc. 
        \item The paper should discuss whether and how consent was obtained from people whose asset is used.
        \item At submission time, remember to anonymize your assets (if applicable). You can either create an anonymized URL or include an anonymized zip file.
    \end{itemize}

\item {\bf Crowdsourcing and Research with Human Subjects}
    \item[] Question: For crowdsourcing experiments and research with human subjects, does the paper include the full text of instructions given to participants and screenshots, if applicable, as well as details about compensation (if any)? 
    \item[] Answer: \answerNA{} 
    \item[] Justification: Our research does not involve crowdsourcing or research with human subjects.
    \item[] Guidelines:
    \begin{itemize}
        \item The answer NA means that the paper does not involve crowdsourcing nor research with human subjects.
        \item Including this information in the supplemental material is fine, but if the main contribution of the paper involves human subjects, then as much detail as possible should be included in the main paper. 
        \item According to the NeurIPS Code of Ethics, workers involved in data collection, curation, or other labor should be paid at least the minimum wage in the country of the data collector. 
    \end{itemize}

\item {\bf Institutional Review Board (IRB) Approvals or Equivalent for Research with Human Subjects}
    \item[] Question: Does the paper describe potential risks incurred by study participants, whether such risks were disclosed to the subjects, and whether Institutional Review Board (IRB) approvals (or an equivalent approval/review based on the requirements of your country or institution) were obtained?
    \item[] Answer: \answerNA{} 
    \item[] Justification: Our research does not involve crowdsourcing or research with human subjects.
    \item[] Guidelines:
    \begin{itemize}
        \item The answer NA means that the paper does not involve crowdsourcing nor research with human subjects.
        \item Depending on the country in which research is conducted, IRB approval (or equivalent) may be required for any human subjects research. If you obtained IRB approval, you should clearly state this in the paper. 
        \item We recognize that the procedures for this may vary significantly between institutions and locations, and we expect authors to adhere to the NeurIPS Code of Ethics and the guidelines for their institution. 
        \item For initial submissions, do not include any information that would break anonymity (if applicable), such as the institution conducting the review.
    \end{itemize}

\end{enumerate}

\end{document}